\documentclass[sigplan,nonacm]{acmart}
\usepackage{booktabs}        
\usepackage{multirow}   
\usepackage{tabularx}
\usepackage{array}
\usepackage{bm}
                                         
\begin{document}
\title{OverlapAdvisor: Modeling and Adapting Computation--Communication Overlap for Distributed MoE Serving}                                           
\author{Chutong Ding}
\affiliation{\institution{Shanghai Jiao Tong University}\country{China}}
\email{dingct2004@sjtu.edu.cn}

\author{Kaixuan Zhang}
\affiliation{\institution{Shanghai Jiao Tong University}\country{China}}
\email{zks1anx@sjtu.edu.cn}

\author{Luping Wang}
\affiliation{\institution{Alibaba Group}\country{China}}
\email{chamu.wlp@alibaba-inc.com}

\author{Shiyou Qian}
\affiliation{\institution{Shanghai Jiao Tong University}\country{China}}
\email{qshiyou@sjtu.edu.cn}

\author{Jian Cao}
\affiliation{\institution{Shanghai Jiaotong University}\country{China}}
\email{cao-jian@sjtu.edu.cn}

\author{Cheng Huang}
\affiliation{\institution{Alibaba Group}\country{China}}
\email{xiaoluo.hc@alibaba-inc.com}

\author{Guodong Yang}
\affiliation{\institution{Alibaba Group}\country{China}}
\email{luren.ygd@taobao.com}

\author{Liping Zhang}
\affiliation{\institution{Alibaba Group}\country{China}}
\email{liping.z@alibaba-inc.com}

\begin{abstract}

Mixture-of-Experts (MoE) inference with expert parallelism (EP) relies on All-to-All communication, making computation--communication overlap strategies critical for achieving high performance. However, no strategy is universally optimal. Across 103 production configurations, varying input lengths, and four hardware platforms, we observe that the fastest strategy shifts based on model shape, EP degree, input length, and hardware. Since no strategy dominates more than $80.6\%$ of configurations, configuration-aware selection is necessary.
In this work, we propose OverlapAdvisor, which models overlap stages using two granularities to capture computation--communication dependencies. OverlapAdvisor builds a unified performance model, accounting for deviations from peak compute, memory, and communication performance across workloads and hardware. This model enables launch-time, configuration-aware strategy selection within two serving engines. We conduct experiments on four MoE models (30B–235B) against three baselines, utilizing SGLang and vLLM. At the kernel level, OverlapAdvisor achieves $\bm{1.16\times}$ average and $\bm{1.59\times}$ maximum speedups, with a prediction MAPE of $\bm{3.3\%}$--$\bm{15.9\%}$. In SGLang and vLLM, it improves end-to-end throughput up to $\bm{31.7\%}$. Its granularity-guided optimizations further achieve median speedups of $\bm{1.10\times}$ for DeepEP-v1 and $\bm{1.22\times}$ for FlashMoE.

\end{abstract}

\maketitle 

\section{Introduction}

Trillion-parameter LLMs~\cite{kimiteam2026kimik2openagentic} exceed single-GPU memory and compute limits, necessitating distributed inference~\cite{kwon2023efficientmemorymanagementlarge,zheng2024sglangefficientexecutionstructured,trifan2025eliminatingmultigpuperformancetaxes}. Mixture-of-Experts (MoE) is a dominant architecture for scaling capacity at fixed compute cost, with expert parallelism (EP)~\cite{lepikhin2020gshardscalinggiantmodels} as its most common distribution strategy. EP routes tokens to experts via All-to-All communication, whose cost scales with token count, hidden dimension, and parallelism degree, creating a critical inference bottleneck~\cite{zheng2025tilelinkgeneratingefficientcomputecommunication,li2026dwdpdistributedweightdata}. Computation--communication overlap is the key to mitigating this bottleneck~\cite{zheng2025tilelinkgeneratingefficientcomputecommunication}.

Many MoE EP strategies use kernel fusion and device-side scheduling for fine-grained communication--computation overlap, reducing launch and synchronization overheads. For example, Comet~\cite{zhang2025cometfinegrainedcomputationcommunicationoverlapping} manages dependencies for fine-grained overlap, while FlashMoE~\cite{aimuyo2025flashmoefastdistributedmoe} fuses overlap stages using device-initiated primitives. Section~\ref{sec:bg-strategies} reviews these and other strategies.

Despite these advances, strategies are typically evaluated on limited model configurations and hardware, leaving no systematic comparison of major strategies under unified conditions. Prior work also implicitly assumes a universally optimal strategy, yet our measurements reveal distinct advantage regions across model shapes, input lengths, EP degrees, and hardware, with no strategy consistently dominating. Context-aware strategy adaptation is thus essential, but existing performance models characterize overlap in strategy-specific ways, preventing a common analytical comparison. Practitioners deploying MoE in SGLang~\cite{zheng2024sglangefficientexecutionstructured} or vLLM~\cite{kwon2023efficientmemorymanagementlarge} must therefore rely on ad-hoc benchmarking or a fixed default strategy.

We propose OverlapAdvisor, a unified model for launch-time strategy adaptation, addressing three challenges. (i) Strategies differ in kernel structure (separate kernels or a megakernel), communication/memory mechanisms (copy engine, TMA, LD/ST), and code implementations (CUTLASS, cuBLASDx, Triton, PTX), making accurate ranking under one unified model difficult. (ii) Achieved bandwidth and throughput deviate substantially from hardware peaks; these deviations vary across the token-count range from decode to prefill and grow with finer work partitioning. (iii) Adaptation must integrate easily into serving engines and cost far less than benchmarking candidates.

Rather than introducing another overlap implementation, OverlapAdvisor treats overlap strategy adaptation as an optimization problem. It models why strategies excel in different regimes to recommend the optimal strategy for any deployment configuration---MoE shape, input length, and EP degree.
To achieve this objective, OverlapAdvisor builds a unified performance model that enables launch-time strategy selection. This model can also analyze the performance bottleneck of existing strategies to suggest optimization solutions. Furthermore, OverlapAdvisor is lightweight and non-intrusive,  
taking about 6 µs without online profiling.

We evaluate OverlapAdvisor against six fixed-strategy baselines, $103$ production MoE shapes, $10$ token counts (from decode to $16$K-token chunked prefill), and four GPU platforms spanning three generations: A100, H20, H800, and GB200. (1) Across 20 strategy--hardware pairs, the performance model achieves $3.3\%$--$15.9\%$ MAPE, and the selector achieves $85.3\%$--$91.6\%$ top-1 accuracy ($88.43\%$ on average). (2) OverlapAdvisor delivers $1.16\times$ average and $1.59\times$ maximum kernel speedups over each platform's strongest fixed strategy, and up to $5.62\times$ over baselines. Even where MegaMoE~\cite{deepgemm2026megamoe} is often optimal, OverlapAdvisor outperforms it by over $5\%$ in up to $23.83\%$ of configurations. (3) In SGLang and vLLM, OverlapAdvisor improves end-to-end throughput over MegaMoE, UniEP~\cite{zheng2026uniepunifiedexpertparallelmoe}, and DeepEP~\cite{deepep2025} by $1.78\%$, $4.57\%$, and $9.97\%$ on average, respectively, with gains up to $31.72\%$. (4) OverlapAdvisor diagnoses granularity bottlenecks and guides finer-grained Dispatch in DeepEP-v1 and coarser-grained Combine in FlashMoE, achieving median speedups of $1.10\times$ and $1.22\times$, respectively.

We make the following contributions:

\begin{itemize}
    \item \textbf{A comprehensive evaluation of MoE EP overlap strategies.} We compare six strategies, including Comet~\cite{zhang2025cometfinegrainedcomputationcommunicationoverlapping}, FlashMoE~\cite{aimuyo2025flashmoefastdistributedmoe}, Torch~\cite{paszke2019pytorchimperativestylehighperformance}, MegaMoE~\cite{deepgemm2026megamoe}, UniEP~\cite{zheng2026uniepunifiedexpertparallelmoe}, and DeepEP~\cite{deepep2025}, under identical conditions across $103$ production MoE shapes, EP $2$--$8$, $10$ input lengths from decode to $16$K prefill, and four platforms spanning three GPU generations.
    \item \textbf{A unified performance model.} Building on existing strategies, we propose a unified performance model that combines granularity-aware analytical modeling for core dependencies with data-driven efficiency modeling for specific configurations. 
    \item \textbf{Implementation and evaluation in mainstream engines.} Implementing OverlapAdvisor within SGLang and vLLM yields significant kernel and end-to-end performance improvements. We further leverage OverlapAdvisor to analyze and optimize existing strategies.
\end{itemize}

\section{Background}
\label{sec:background}

\subsection{MoE and EP}
\label{sec:bg-moe}

An MoE layer replaces dense feed-forward networks with multiple experts, activating only the top-$k$ per token to scale capacity without increasing compute cost. EP~\cite{lepikhin2020gshardscalinggiantmodels} distributes these experts across GPUs. Dispatch routes tokens to experts for computation, while Combine returns outputs to their origins. Both rely on All-to-All communication, whose cost depends on token count, hidden dimension, and EP degree, often creating a bottleneck in MoE inference~\cite{zheng2025tilelinkgeneratingefficientcomputecommunication,li2026dwdpdistributedweightdata}.

\subsection{Overlap Strategies for MoE EP}
\label{sec:bg-strategies}

Computation--communication overlap enables concurrent execution and data transfer across streaming multiprocessors (SMs), copy engines, or dedicated warps. An \emph{overlap strategy} partitions and schedules MoE operations to exploit this concurrency.
These strategies integrate communication and computation via device-side scheduling to minimize launch and synchronization overhead. For example, Comet~\cite{zhang2025cometfinegrainedcomputationcommunicationoverlapping} resolves dependencies for fine-grained overlap; FlashMoE~\cite{aimuyo2025flashmoefastdistributedmoe} utilizes device-initiated primitives to fuse overlap stages; and UniEP~\cite{zheng2026uniepunifiedexpertparallelmoe} unifies strategies through parameterized MegaKernel abstractions. We evaluate Comet, FlashMoE, MegaMoE, and UniEP as overlap strategies, while using Torch- and DeepEP-based implementations as serialized baselines with sequential communication and computation. The DeepEP-based baseline uses DeepEP for communication and DeepGEMM~\cite{deepgemm2025} for computation.

Strategies differ in kernel partitioning, execution mechanisms, and code implementations.
(1) \emph{Communication mechanisms} transfer data across ranks via load/store pipeline (LD/ST), Tensor Memory Accelerator (TMA), or copy engines (CEs).
(2) \emph{Memory mechanisms} load data from local high bandwidth memory (HBM) via LD/ST or TMA. Both communication and memory loading use LD/ST, but with different instructions and transfer semantics: the former accesses peer memory via \texttt{ld.global}/\texttt{st.global}, while the latter uses \texttt{cp.async} for asynchronous local HBM-to-shared-memory transfers that bypass registers.
(3) \emph{Computation mechanisms} use Tensor Cores for expert computation and CUDA Cores for activations and other element-wise operations.
(4) \emph{Implementations} realize these mechanisms using low-level instructions (CUDA C++ with inline PTX), GPU DSLs (Triton), or CUDA libraries (CUTLASS and cuBLASDx).

Mechanism choices and their code implementations affect instruction scheduling and resource efficiency, yielding performance differences among similar strategies. Table~\ref{tab:mechanism} summarizes the execution mechanisms and implementations of the evaluated strategies.

\begin{table}[tbp]
    \centering
    \caption{Execution mechanisms and implementations of MoE EP strategies.}
    \vspace{-3mm}
    \label{tab:mechanism}
    \footnotesize
    \setlength{\tabcolsep}{3pt}
    \renewcommand{\arraystretch}{1.05}
    \begin{tabular}{@{}lccccl@{}}
        \toprule
        \multirow{2}{*}{\textbf{Strategy}}
          & \multicolumn{2}{c}{\textbf{Comm}}
          & \multirow{2}{*}{\textbf{Mem}}
          & \multirow{2}{*}{\textbf{Comp}}
          & \multirow{2}{*}{\textbf{Impl}} \\
        \cmidrule(lr){2-3}
          & Dispatch & Combine & & & \\
        \midrule
        Comet
          & CE & LD/ST & LD/ST, TMA & Tensor Core & CUTLASS \\
        FlashMoE
          & LD/ST & LD/ST & LD/ST & Tensor Core & cuBLASDx \\
        UniEP
          & TMA & LD/ST & TMA & Tensor Core & Triton \\
        MegaMoE
          & TMA & LD/ST & TMA & Tensor Core & CUDA C++ \\
        DeepEP-v1
          & LD/ST & LD/ST
          & \multirow{2}{*}{TMA}
          & \multirow{2}{*}{Tensor Core}
          & \multirow{2}{*}{CUDA C++} \\
        DeepEP-v2
          & TMA & TMA & & & \\
        \bottomrule
    \end{tabular}
    \vspace{-3mm}
\end{table}

\subsection{Strategy Selection and Modeling}

\begin{figure}[!htbp]
\centering
\includegraphics[width=0.48\textwidth]{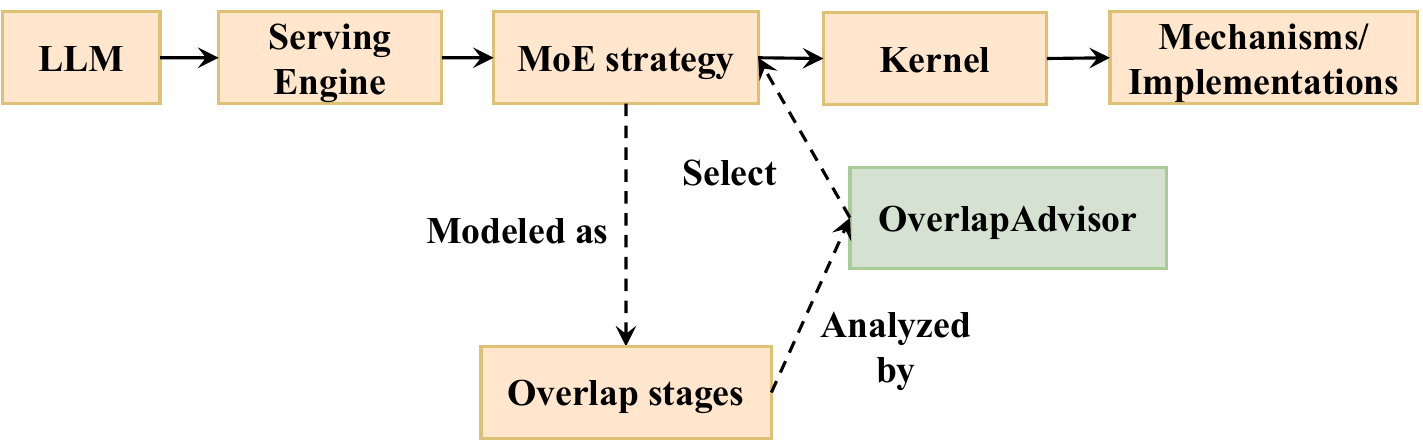}
\vspace{-6mm}
\caption{OverlapAdvisor in the execution stack. Solid arrows show execution flow; dashed arrows show OverlapAdvisor's optimal strategy selection via two-stage overlap analysis.}
\label{fig:background}
\vspace{-4mm}
\end{figure}

Figure~\ref{fig:background} locates OverlapAdvisor in the execution stack: a serving engine runs an LLM whose MoE layers use EP strategies implemented by kernels with specific execution mechanisms and implementations. Given the router's assignments, OverlapAdvisor models each strategy's full \emph{MoE layer} from Dispatch to Combine through two logical overlap stages in the execution pipeline: Dispatch+GroupGEMM couples dispatch with the first GroupGEMM, while GroupGEMM+Combine couples the second GroupGEMM with Combine. These stages are analytical units, not kernel boundaries: a strategy may split a stage across multiple kernels or fuse both stages into one kernel. OverlapAdvisor uses this stage-based analysis to predict each strategy's MoE execution time and select the fastest strategy.

\subsection{Key Notations and Workload Configuration}
\label{sec:bg-config}

Table~\ref{tab:Notations} lists the key notations used in this paper, covering workload dimensions, hardware capabilities, kernel settings, and performance-model quantities.
We define a \emph{workload configuration} as the tuple $(H, I, E, k, S_{R_{\mathrm{in}}}, D)$, capturing model dimensions, token counts, and parallelism settings, independent of hardware and strategy.

\begin{table}[!htbp]
\centering
\caption{Notations used in this work.}
\vspace{-3mm}
\label{tab:Notations}
\footnotesize
\setlength{\tabcolsep}{4pt}
\renewcommand{\arraystretch}{1.05}
\begin{tabular}{@{}ll@{}}
\toprule
\textbf{Symbol} & \textbf{Description} \\
\midrule
\multicolumn{2}{@{}l}{\textbf{Workload}} \\
$H$, $I$ & Hidden size, MoE intermediate size \\
$E$ & Total number of experts \\
$k$ & Number of experts per token (top-$k$) \\
$S_{R_{\mathrm{in}}}$ & Input tokens per rank \\
$D$ & Expert-parallelism degree \\
$\beta$ & Bytes per element (BF16/FP8) \\
$S_{R_{\mathrm{a}}} = S_{R_{\mathrm{in}}} \cdot k$
    & Tokens assigned per rank (average) \\
$S^{\max}_{R_{\mathrm{a}}}$
    & Tokens assigned to the hottest rank \\
$\rho = S^{\max}_{R_{\mathrm{a}}}/S_{R_{\mathrm{a}}}$
    & Imbalance factor \\
$E_{\mathrm{L}} = E/D$
    & Local experts per rank \\
$S_E = S_{R_{\mathrm{a}}}/E_{\mathrm{L}}$
    & Tokens assigned per expert (average) \\
\addlinespace[2pt]
\multicolumn{2}{@{}l}{\textbf{Hardware}} \\
$F$ & Peak compute rate (BF16/FP8) \\
$B_{\mathrm{HBM}}$, $B_{\mathrm{NVL}}$ & Peak HBM, NVLink bandwidth \\
$N_{\mathrm{SM}}$ & Number of SMs on the GPU \\
\addlinespace[2pt]
\multicolumn{2}{@{}l}{\textbf{Kernel}} \\
$B_M$, $B_N$ & Block M and block N in GroupGEMM \\
\addlinespace[2pt]
\multicolumn{2}{@{}l}{\textbf{Performance Model}} \\
$T$ & Total MoE execution time \\
$T_{\mathrm{comp}}$, $T_{\mathrm{comm}}$
    & Computation / communication time \\
$G_{\mathrm{comp}}$, $G_{\mathrm{comm}}$
    & Computation / communication granularity \\
\bottomrule
\end{tabular}
\vspace{-4mm}
\end{table}

\section{Observations}
\label{sec:motivation}

\begin{figure}[!htbp]
\centering
\includegraphics[width=0.48\textwidth]{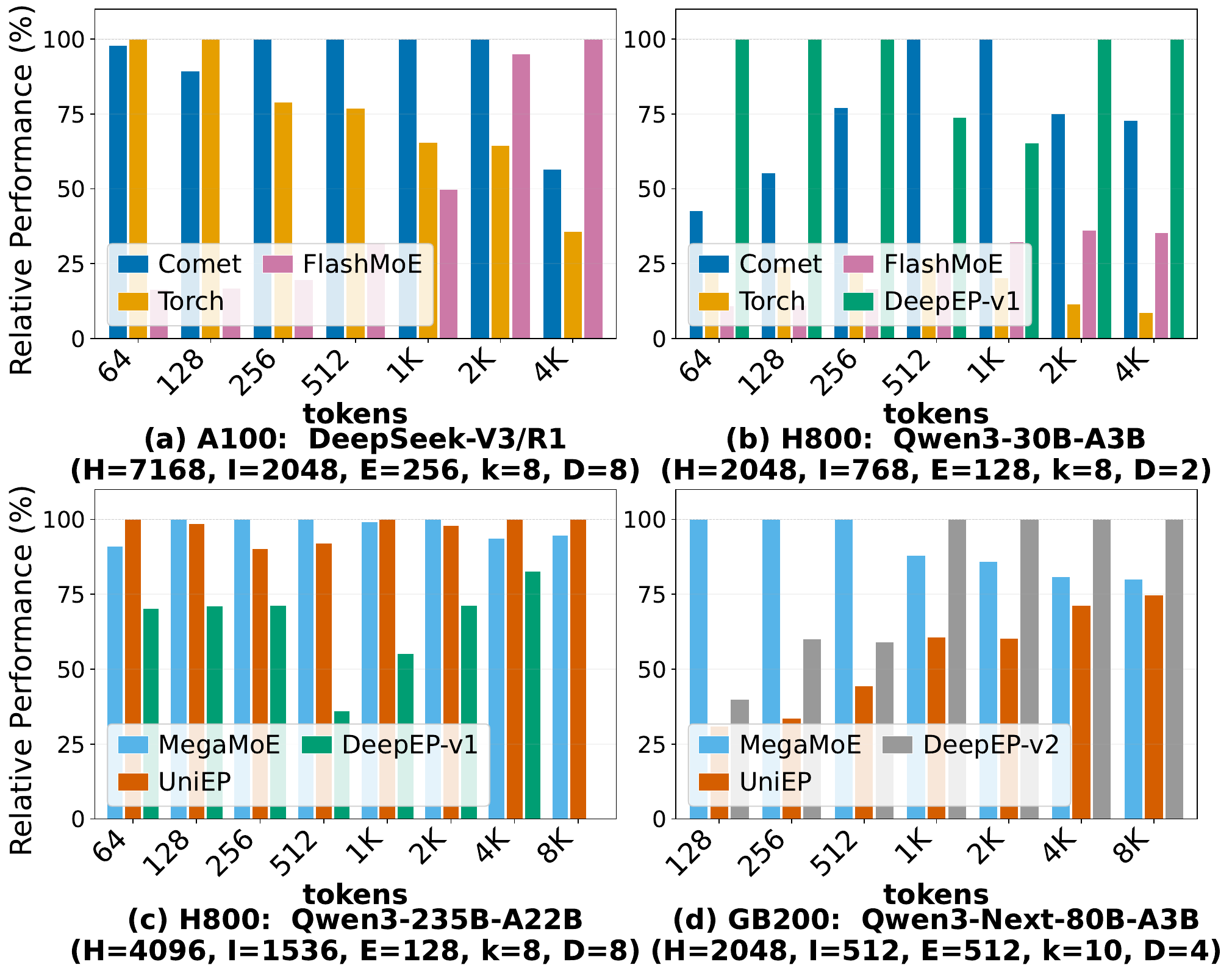}
\vspace{-7mm}
\caption{Relative performance of six MoE EP strategies at the kernel level, normalized per configuration to the fastest strategy (best is $100\%$). (a) and (b): BF16 with a standard MLP; (c) and (d): FP8 with a gated MLP.}
\label{fig:introduction}
\vspace{-3mm}
\end{figure}

MoE EP strategy performance depends on various factors: the communication mechanism, token layout, resource partitioning, synchronization boundaries, and hardware characteristics such as SM count and interconnect bandwidth. Fixing these choices limits a strategy's effectiveness across deployments, yet prior work often assumes universal optimality. Figure~\ref{fig:introduction} compares six MoE EP strategies, using single-layer kernel-level benchmarks with MoE configurations from DeepSeek-V3~\cite{deepseekai2025deepseekv3technicalreport} and Qwen3~\cite{yang2025qwen3technicalreport} on A100, H800 and GB200. The evaluation covers two settings: BF16 with a standard MLP and FP8 with a gated MLP. No single strategy is optimal across the evaluated configurations and hardware platforms; different strategies excel in different regions. From this analysis, we identify three observations.

\noindent\textbf{Observation 1: Existing strategies lack general effectiveness.}
\label{sec:mot-eval-gap}

Figure~\ref{fig:introduction} shows that different strategies excel under different conditions, but prior evaluations cover limited configurations and may miss these trade-offs. 
Table~\ref{tab:eval-scope-small} summarizes the evaluation scope of strategies by the number of reported $(H,I,E,k,S_{R_{\mathrm{in}}},D)$ configurations. We omit $D$ as a separate column because it remains fixed for each $(H,I,E,k,S_{R_{\mathrm{in}}})$ tuple. Existing evaluations utilize diverse, self-selected hardware and workloads. For instance, FlashMoE compares against Comet and Megatron~\cite{narayanan2021efficientlargescalelanguagemodel} on H100 across 14 configurations, while Comet compares against Megatron and FasterMoE~\cite{yang2025fastermoellminference} on H800 and L20 across 22 configurations. Due to the lack of a common platform or unified configuration, these results neither facilitate direct comparison nor establish general performance metrics beyond the specific workloads evaluated. 

\begin{table}[bp]
\vspace{-3mm}
\centering
\caption{Evaluation scope of MoE overlapping strategies.}
\vspace{-3mm}
\label{tab:eval-scope-small}
\footnotesize
\setlength{\tabcolsep}{2pt}
\renewcommand{\arraystretch}{1.05}
\begin{tabularx}{\columnwidth}{@{}lccc
    >{\raggedright\arraybackslash}X@{}}
\toprule
Strategy
& \textbf{$(H,I,E,k)$}
& \textbf{$S_{R_{\mathrm{in}}}$}
& Configurations
& Platforms \\
\midrule
Comet    & 10 & 5 & 22 & H800, L20 \\
FlashMoE & 6  & 3 & 14 & H100 \\
UniEP    & 12 & 2 & 24 & 2 Hopper clusters \\
\midrule
\textbf{OverlapAdvisor}
& \textbf{103}
& \textbf{10}
& \textbf{1030}
& \textbf{A100, H20, H800, GB200} \\
\bottomrule
\end{tabularx}
\end{table}

\noindent\textbf{Observation 2: Existing kernel performance models are strategy-specific, not unified.}
\label{sec:mot-model-gap}

Current kernel performance models fail to support context-aware strategy selection. Most models approximate overlap as $\max(T_{\mathrm{comp}},T_{\mathrm{comm}})$, while refinements involving non-overlappable terms remain strategy-specific~\cite{sul2025parallelkittenssystematicpracticalsimplification,zheng2026uniepunifiedexpertparallelmoe}. Lacking a unified performance model for guidance, practitioners rely on costly benchmarking or potentially suboptimal defaults.

\noindent\textbf{Observation 3: Effective strategies require configuration adaptation.}
\label{sec:mot-context}

Achieving effective MoE EP overlap is challenging, which is manifested in three aspects. First, input-dependent routing creates dynamic communication--computation dependencies, which complicates static partitioning for fine-grained pipelining~\cite{zhang2025cometfinegrainedcomputationcommunicationoverlapping}. Second, mapping tokens to destination buffers introduces extra metadata and synchronization costs~\cite{zhou2026moehubtamingsoftwarecomplexity}. Third, overlap also causes contention for SM resources, memory bandwidth, and L2 cache~\cite{cui2026resourceawarecomputationcommunicationoverlapmultigpu,zhu2025nanoflowoptimallargelanguage}. Furthermore, GPU multi-stream scheduling remains non-deterministic in real-world production environments~\cite{chang2024fluxfastsoftwarebasedcommunication}. Consequently, selecting the right strategy requires balancing both hardware and workload factors, resulting in a complex, high-dimensional optimization problem that makes manual tuning impractical. 

\section{Design}
\label{sec:Design}
\subsection{Overview}
\begin{figure}[htbp] 
\centering 
\includegraphics[width=0.48\textwidth]{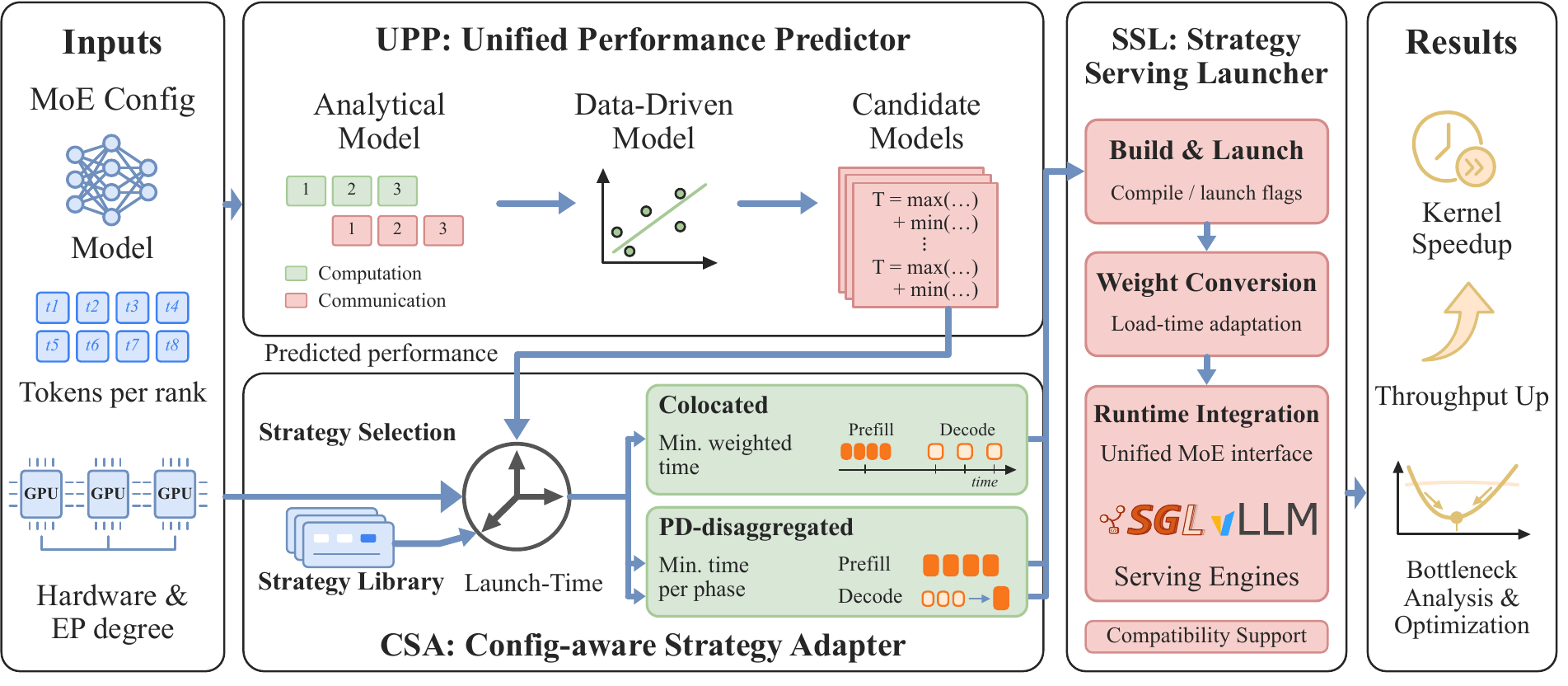}
\caption{Overall structure of OverlapAdvisor.}
\label{fig:design1} 
\end{figure}

As illustrated in Figure~\ref{fig:design1}, the overall structure of OverlapAdvisor consists of three components.

\textbf{A Unified Performance Predictor (UPP).} To enhance both generality and accuracy, we employ a hybrid approach to model the kernel performance of both overlap and serial strategies. 
First, OverlapAdvisor constructs an analytical model for MoE strategies from the perspective of execution granularity. 
Second, we treat efficiency-related parameters that cannot be directly derived from hardware specifications, such as memory and communication bandwidth utilization and peak computational efficiency, as strategy-specific variables to be fitted. 
OverlapAdvisor samples the execution time of each strategy using only a limited set of configurations, and utilizes a differential evolution (DE)~\cite{Storn1997DifferentialE} optimizer to estimate these parameters. 
Finally, incorporating these fitted parameters yields hardware- and strategy-aware models that predict performance for arbitrary configurations (\S\ref{sec:UPP}).

\textbf{A Configuration-aware Strategy Adapter (CSA).} At server launch, OverlapAdvisor reads the deployment configuration and uses UPP to select strategies that minimize predicted MoE execution time. For PD-disaggregated deployments, CSA selects prefill and decode strategies independently; for colocated deployments, it selects a shared strategy by weighting each phase's predicted time. Since all inputs are deployment invariants, selection runs once in about $6$~$\mu$s, requiring no profiling or runtime adaptation (\S\ref{sec:Adapter}).

\textbf{A Strategy Serving Launcher (SSL).} Since a decision is only valuable if it can be implemented, multiple strategies are integrated into serving engines (such as SGLang~\cite{zheng2024sglangefficientexecutionstructured}, vLLM~\cite{kwon2023efficientmemorymanagementlarge}) via a unified, interchangeable MoE backend interface. SSL applies CSA's selections through launch options or environment variables, supporting hardware-specific kernel builds, load-time weight conversion, and lightweight runtime integration. These additive changes require only minor compatibility adjustments, preserve existing code paths, and leave non-MoE layers unchanged (\S\ref{sec:strategy-integrator}).

\subsection{Unified Performance Predictor (UPP)}
\label{sec:UPP}
Predicting a strategy's kernel execution time requires two key components: an analytical model of the fundamental computation--communication dependencies based on a granularity abstraction, independent of strategy-specific kernel implementations; and data-driven fitting of efficiency parameters that capture bandwidth utilization and compute efficiency while retaining a shared prediction structure across strategies. OverlapAdvisor derives the former analytically (\S\ref{sec:analytical}) and fits the latter from a small sample set (\S\ref{sec:data-driven}).

\subsubsection{Granularity-aware Analytical Modeling}
\label{sec:analytical}
We focus on Dispatch + GroupGEMM and GroupGEMM + Combine, which dominate MoE layer execution time and capture key overlap differences across strategies. The total execution time is determined by combining these two stages as follows:
\begin{equation}
  \widehat{T} = T^{(1)} + T^{(2)} + \sum_{j} T_{j}^{\mathrm{misc}},
\label{eq:layer}
\end{equation}
where $T^{(1)}$ and $T^{(2)}$ model Dispatch+GroupGEMM and Group\\GEMM+Combine, respectively. $T_j^{\mathrm{misc}}$ covers memory-bound kernels (e.g., SwiGLU/GELU element-wise activations, permute that groups tokens by expert, and unpermute that restores their original order).

Modeling the two stages involves characterizing two factors: the granularity at which communication data enables computation, and the granularity at which computation results enable communication.
Let \(T_{\mathrm{comp}}\) and \(T_{\mathrm{comm}}\) represent the computation and communication time within the Dispatch+GroupGEMM or GroupGEMM+Combine stages. Serialized execution gives $T=T_{\mathrm{comp}}+T_{\mathrm{comm}}$, reflecting strict data dependencies; perfect overlap gives $T=\max(T_{\mathrm{comp}},T_{\mathrm{comm}})$, assuming no such dependencies. In practice, however, the execution time falls between these two bounds because computation and communication exhibit data dependencies at a specific granularity. 
We abstract each computation-- communication stage using two granularities, $G_{\mathrm{comp}}$ and $G_{\mathrm{comm}}$, defined as the numbers of rounds into which computation and communication are partitioned. These quantities characterize the granularity of data dependencies between computation and communication.

\begin{figure}[bp] 
\vspace{-4mm}
\centering 
\includegraphics[width=0.48\textwidth]{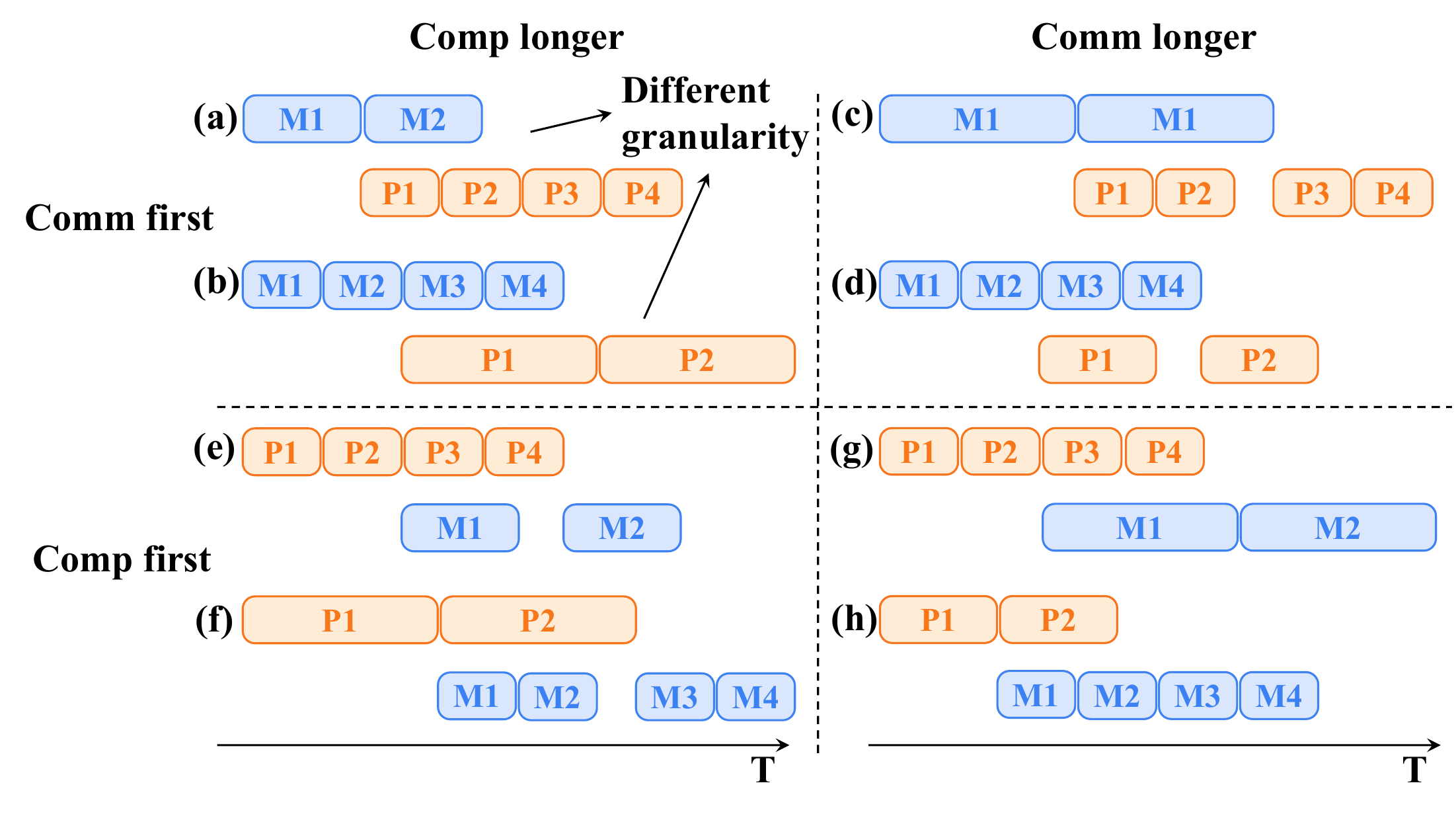}
\vspace{-8mm}
\caption{Overlap schematic (M: communication; P: computation), showing how $T$ is determined by the sequence of execution--whether computation or communication takes precedence--as well as the dominant time factor and the coarser granularity between the two.} 
\label{fig:design2} 
\end{figure}

Overall, in each MoE EP stage, three factors impact execution time. 
The first is the temporal ordering of computation and communication. 
The second is the execution time of computation and communication. As the communication workload is \( O(S_{R_{\mathrm{in}}} \cdot H) \) while the computation workload is \( O(S_{R_{\mathrm{in}}} \cdot H \cdot I) \), the dominant term alternates between \(T_{\mathrm{comp}}\) and \(T_{\mathrm{comm}}\) as $I$ varies. 
The third is the granularity, which partitions tasks into various rounds.
Figure~\ref{fig:design2} shows stage execution times for all eight combinations of the three factors. 
We model each computation--communication stage as:
\begin{equation}
    T = \max(T_{\mathrm{comp}}, T_{\mathrm{comm}})
    + \frac{\min(T_{\mathrm{comp}}, T_{\mathrm{comm}})}
    {\min(G_{\mathrm{comp}}, G_{\mathrm{comm}})},
\label{eq:overlap-time}
\end{equation}
For example, Figure~\ref{fig:design2}(a) has $T_{\mathrm{comp}}>T_{\mathrm{comm}}$, $G_{\mathrm{comp}}=4$, and $G_{\mathrm{comm}}=2$, giving $T=T_{\mathrm{comp}}+T_{\mathrm{comm}}/2$. 

Contention for SM resources, L2 cache, and bandwidth during overlap is captured by the data-driven model below rather than explicitly modeled in this equation.

\subsubsection{Data-Driven Efficiency Modeling for Strategies and Platforms}
\label{sec:data-driven}
Routing-induced load imbalance across ranks affects MoE execution time. Following~\cite{deepep_pr393}, we quantify it as $\rho=S^{\max}_{R_{\mathrm{a}}}/S_{R_{\mathrm{a}}}$, where $S_{R_{\mathrm{a}}}$ and $S^{\max}_{R_{\mathrm{a}}}$ denote the average and maximum numbers of tokens received per rank, respectively; $\rho=1$ indicates uniform loading. To model inter-rank imbalance, we approximate the load of each local expert on the busiest rank by the rank's average expert load, $\rho S_E = S^{\max}_{R_{\mathrm{a}}}/E_{\mathrm{L}}$, where $E_{\mathrm{L}}$ is the number of local experts per rank and $S_E = S_{R_{\mathrm{a}}}/E_{\mathrm{L}}$ is the average expert load. This approximation does not explicitly capture load imbalance among experts within a rank. We evaluate all granularities at this busiest-rank load to approximate the execution bottleneck.

While the analytical model captures the overlap structure, it fails to fully characterize actual performance. Data-driven efficiency modeling improves accuracy and distinguishes between strategies with similar overlap granularities but different efficiencies, such as MegaMoE and UniEP. The two primary factors affecting resource efficiency in MoE EP are bandwidth and GEMM.

\textbf{Bandwidth Efficiency Modeling.} In GPU systems, the effective bandwidth of NVLink and HBM varies with the volume of data transferred~\cite{sul2025parallelkittenssystematicpracticalsimplification}. At small volumes, startup overheads, such as kernel launch or communication startup, account for a large fraction of execution time, and limited parallelism further constrains bandwidth utilization. As data volume increases, these overheads are amortized, and effective bandwidth generally rises before reaching saturation. The saturation level can be below the hardware peak and depends on the platform and the data-movement mechanism, such as CE, TMA, or LD/ST instructions (Table~\ref{tab:mechanism}).

UPP uses a shared capped power law to model communication and HBM bandwidth efficiency across strategies:
\begin{equation}
    \eta_{\mathrm{comm}/\mathrm{mem}} = \min\left(\eta_{cap}, \left(\frac{V}{V_{ref}}\right)^{\alpha}\right),
\label{eq:bandwidth-efficiency}
\end{equation}
where $\eta_{\mathrm{cap}}$ is the saturation efficiency, $V$ is the data volume, $V_{\mathrm{ref}}$ is a reference volume that sets the scale of the efficiency curve, and $\alpha$ is the power-law exponent. 
Below $V_{\mathrm{sat}}=V_{\mathrm{ref}}\eta_{\mathrm{cap}}^{1/\alpha}$, efficiency increases with $V$, and transfer time grows as $V^{1-\alpha}$. Above saturation, efficiency stays at $\eta_{\mathrm{cap}}$ and transfer time grows linearly. For a fixed saturation point, larger $\alpha$ gives a steeper efficiency decline at smaller volumes. We fit $\eta_{\mathrm{cap}}$, $V_{\mathrm{ref}}$, and $\alpha$ for each strategy and platform.

We model two primary communication topologies: ring-based (Comet dispatch; Torch) and P2P (Comet combine; FlashMoE, MegaMoE, UniEP, DeepEP), as follows:
\begin{align}
V_{\mathrm{comm}} &= \begin{cases}
        (D-1) \cdot S_{R_{\mathrm{in}}} \cdot H \cdot \beta, & \text{ring-based}, \\[6pt]
        \rho \cdot (D-1) \cdot \dfrac{S_{R_{\mathrm{in}}}}{D} \cdot k \cdot H \cdot \beta, & \text{P2P-based}.
     \end{cases} \\[10pt]
T_{\mathrm{comm}} &= \frac{V_{\mathrm{comm}}}{B_{\mathrm{NVL}} \cdot \eta_{\mathrm{comm}}(V_{\mathrm{comm}}/G_{\mathrm{comm}})},
\end{align}
where $D$ is the EP degree, $S_{R_{\mathrm{in}}}$ the input tokens per rank, $H$ the hidden size, $k$ selected experts, $\beta$ the bytes per element (e.g., BF16 or FP8), and $B_{\mathrm{NVL}}$ the peak NVLink bandwidth. P2P communication time is determined by the busiest rank's receive load, and $\eta_{\mathrm{comm}}$ is evaluated at the per-round volume $V_{\mathrm{comm}}/G_{\mathrm{comm}}$. 

GroupGEMM memory-access time $T_{\mathrm{mem}}$ is modeled as:
\begin{align}
V_{\mathrm{mem}}^{(i)} &=
\begin{cases}
S^{\max}_{R_{\mathrm{a}}}\,(H + m I)\,\beta \;+\; E_{\mathrm{L}} \cdot m \cdot H I \beta, & i = 1,\\[3pt]
S^{\max}_{R_{\mathrm{a}}}\,(I + H)\,\beta \;+\; E_{\mathrm{L}} \cdot H I \beta, & i = 2,
\end{cases} \\[3pt]
T_{\mathrm{mem}}^{(i)} &= \frac{V_{\mathrm{mem}}^{(i)}}
   {B_{\mathrm{HBM}} \cdot \eta_{\mathrm{mem}}\!\left(V_{\mathrm{mem}}^{(i)}/G_{\mathrm{comp}}\right)},
\end{align}
where $i=1,2$ denote up- and down-projection GroupGEMMs, respectively, and $m$ denotes the number of first-layer projection matrices ($1$ for standard MLPs, $2$ for gated MLPs). $S^{\max}_{R_{\mathrm{a}}}$ is the busiest rank's received token count, $E_{\mathrm{L}}$ local experts per rank, $I$ the MoE intermediate size, and $B_{\mathrm{HBM}}$ the peak HBM bandwidth. Since each GroupGEMM runs in $G_{\mathrm{comp}}$ rounds, $\eta_{\mathrm{mem}}$ is evaluated at the per-round volume $V_{\mathrm{mem}}^{(i)}/G_{\mathrm{comp}}$. Small memory-bound kernels, including activation kernels, permute, and unpermute, are modeled similarly.

\textbf{GEMM Efficiency Modeling.} Following the Roofline model~\cite{10.1145/1498765.1498785}, GroupGEMM is limited by compute or memory performance. As shown in Table~\ref{tab:mechanism}, all strategies use Tensor Cores but differ in memory-access mechanisms and kernel implementations. The bottleneck depends on problem size: more tokens per expert increase weight reuse and arithmetic intensity, shifting GroupGEMM toward the compute-bound regime. Conversely, when the number of tokens is small, the memory-access overhead of the weight matrices cannot be sufficiently amortized, causing the operation to be memory-bound. Consequently, in the compute-bound regime, a large number of tiles are evenly distributed across all SMs, and computational efficiency remains approximately constant at $\eta_{gemm}$. In the memory-bound regime, memory-access efficiency varies with data volume and is modeled using the same power-law efficiency formulation as Eq.~\eqref{eq:bandwidth-efficiency}. Here, $\eta_{\mathrm{gemm}}$ is fitted for each strategy and hardware.

We model GroupGEMM execution time  $T_{\mathrm{comp}}$ as:
\begin{align}
C_{\mathrm{gemm}}^{(i)} &=
\begin{cases}
2\,m\,S^{\max}_{R_{\mathrm{a}}}\,H I \,\gamma_{\mathrm{pad}}, & i = 1,\\[3pt]
2\,S^{\max}_{R_{\mathrm{a}}}\,H I \,\gamma_{\mathrm{pad}}, & i = 2,
\end{cases} \\[3pt]
T_{\mathrm{comp}}^{(i)} &= \max\!\left(\frac{C_{\mathrm{gemm}}^{(i)}}
   {F \cdot \eta_{\mathrm{gemm}}},\; T_{\mathrm{mem}}^{(i)}\right),
\end{align}
where $C_{\mathrm{gemm}}$ is the floating-point workload of GroupGEMM, $F$ is the peak compute rate (TFLOPs), and $\gamma_{\mathrm{pad}} = \frac{\left\lceil \rho S_E/B_M \right\rceil B_M}{\rho S_E}$ is the padding inflation factor along the token dimension, which accounts for the computational waste incurred when $\rho S_E$ is not an integer multiple of $B_M$.

\textbf{Parameter Fitting.}
The performance models established above for MoE EP strategies include several physical parameters that must be fitted, primarily the bandwidth efficiency parameters ($\eta_{cap}$, $V_{ref}$, $\alpha$) and the GEMM efficiency parameter ($\eta_{\mathrm{gemm}}$). These parameters cannot be read off directly from hardware specifications.
For each $\langle$strategy, platform$\rangle$ pair, we fit the relevant parameters separately for each kernel (e.g., Dispatch, permute, GroupGEMM, and GELU) by minimizing MAPE over $n$ measured configurations. Parameter bounds enforce physical validity (e.g., $\alpha,\eta_{\mathrm{gemm}}\in(0,1)$). Since the model's $\min$, $\max$, and $\lceil\cdot\rceil$ operations make the objective nonsmooth and nonconvex, we use DE for global optimization rather than gradient-based methods~\cite{Liu1989OnTL,kingma2017adammethodstochasticoptimization}.

\subsection{Configuration-aware Strategy Adapter (CSA)}
\label{sec:Adapter}

MoE EP strategies differ in overlap behavior, execution mechanisms and implementations, with no universal winner. Therefore, an effective serving engine must select the strategy that maximizes end-to-end throughput. OverlapAdvisor performs model-driven, launch-time selection: it uses UPP to predict candidate performance from the deployment configuration and selects the best strategy before server startup, without online profiling or runtime adaptation. Strategy selection is phase-aware because prefill and decode operate at significantly different sequence lengths.

With data-parallel attention, prefill and decode have different input token counts. The decode per-kernel token count $S_{R_{\mathrm{in}}}^{\mathrm{dec}}$ equals the number of concurrent requests per rank, as each request contributes one token per step. The prefill per-kernel token count $S_{R_{\mathrm{in}}}^{\mathrm{pre}}$ is set by the per-rank chunked-prefill budget~\cite{agrawal2023sarathiefficientllminference}. For example, 32 decode requests yield $S_{R_{\mathrm{in}}}^{\mathrm{dec}}=32$, versus $S_{R_{\mathrm{in}}}^{\mathrm{pre}}=4096$ for a full 4096-token chunk. These differences place the phases at different points on the efficiency curves in Eq.~\eqref{eq:bandwidth-efficiency}, potentially favoring different backends even for the same model and EP degree.

For each phase $\varphi\in\{\mathrm{pre},\mathrm{dec}\}$, OverlapAdvisor selects
\begin{equation}
  s^{\star}_{\varphi} = \arg\min_{s\in\mathcal{S}}\;\widehat{T}_{s,\varphi}\!\left(H,I,E,k,S_{R_{\mathrm{in}}}^{\varphi},D\right),
\label{eq:select}
\end{equation}
where $\mathcal{S}$ contains the strategies available on the target platform and $\widehat{T}_{s,\varphi}$ is the predicted MoE execution time. All inputs are deployment configurations fixed during server operation, so ranking is computed only once.

CSA's launch-time selection avoids three limitations: (i) fixed strategies are suboptimal as rankings vary across configurations; (ii) profiling-based autotuning benchmarks every candidate on the target cluster at each launch or configuration change, incurring profiling overhead, whereas our model is fitted once per platform; and (iii) runtime adaptation requires online strategy evaluation and may incur switching and warm-up costs. Instead, Eq.~\eqref{eq:select} evaluates $|\mathcal{S}|$ candidates in closed form using a few hundred arithmetic operations per phase, completing selection in approximately $6$~$\mu$s at launch.

OverlapAdvisor supports both PD-disaggregated deployments with separate phase-specific strategies and colocated deployments sharing one backend.

\textbf{PD Disaggregation.} Prefill and decode run on separate GPU sets with EP degrees $D_{\mathrm{p}}$ and $D_{\mathrm{d}}$. OverlapAdvisor applies Eq.~\eqref{eq:select} at $(D_{\mathrm{p}},S_{R_{\mathrm{in}}}^{\mathrm{pre}})$ and $(D_{\mathrm{d}},S_{R_{\mathrm{in}}}^{\mathrm{dec}})$, configuring each instance via separate flags. Unlike fixed-backend systems, this permits heterogeneous choices, such as DeepEP for prefill and MegaMoE for decode.

\textbf{Colocated Deployment.} A shared backend minimizes total MoE time per GPU. For $L$ MoE layers, $B$ concurrent requests per rank, input length $\mathrm{ISL}$, and output length $\mathrm{OSL}$, the instance executes $L B\,\mathrm{ISL}/S_{R_{\mathrm{in}}}^{\mathrm{pre}}$ prefill and $L\,\mathrm{OSL}$ decode MoE invocations. Weighting each phase's predicted time by its invocation count gives
\begin{equation}
  s^{\star} = \arg\min_{s\in\mathcal{S}}\;
  \frac{L \cdot B \cdot \mathrm{ISL}}{S_{R_{\mathrm{in}}}^{\mathrm{pre}}}\,\widehat{T}_{s}\!\left(D,S_{R_{\mathrm{in}}}^{\mathrm{pre}}\right)
  +
  L \cdot \mathrm{OSL}\,\widehat{T}_{s}\!\left(D,S_{R_{\mathrm{in}}}^{\mathrm{dec}}\right).
\label{eq:select-colocated}
\end{equation}

\subsection{Strategy Serving Launcher (SSL)}
\label{sec:strategy-integrator}

To support strategy selection, we integrate candidate strategies as interchangeable MoE backends in SGLang and vLLM, selected via existing launch options or environment variables (e.g., \texttt{--moe-a2a-backend} in SGLang and \texttt{--all2all-backend} in vLLM), leaving the rest of the serving stack unchanged.

Integration comprises three steps and minor compatibility adjustments, all additive and preserving existing code paths:
(S1)~\emph{Build and Launch} ($\approx$0.03k lines per engine): expose kernels through hardware-specific compile flags and add selector launch options.
(S2)~\emph{Weight Conversion}: convert checkpoints at load time to the required kernel layout in SGLang ($\approx$0.15k lines), and adapt weights through native online block-FP8 quantization in vLLM ($\approx$0.05k lines).
(S3)~\emph{Runtime Path}: prepare activations and routing metadata in the backend's cross-rank buffer and invoke its kernels through a lightweight MoE-forward branch in SGLang ($\approx$0.5k lines) or a wrapper at vLLM's MoE-method selection point ($\approx$0.35k lines). Non-MoE layers remain unchanged.
\emph{Compatibility} ($\approx$0.02k lines per engine): bypass unquantized-weight fast paths during online quantization, adapt configurations to newer model-library versions, and size communication buffers for single-node use.

The changes span $\approx$0.7k lines across 11 files in SGLang ($\sim$0.1\% of ${\sim}600$k Python lines) and $\approx$0.4k lines across 3 files in vLLM ($\sim$0.1\% of ${\sim}320$k). Once weight conversion and runtime paths are in place, strategies sharing the same buffer and weight contracts need only minor kernel-call updates, typically tens of lines.

\section{Implementation}
\subsection{Strategy Granularity Instantiation for Analytical Modeling}
Table~\ref{tab:granularity} instantiates the two granularities defined in Section~\ref{sec:analytical} for six representative MoE EP strategies: four overlap approaches (Comet~\cite{zhang2025cometfinegrainedcomputationcommunicationoverlapping}, FlashMoE~\cite{aimuyo2025flashmoefastdistributedmoe}, MegaMoE~\cite{deepgemm2026megamoe}, UniEP~\cite{zheng2026uniepunifiedexpertparallelmoe}) and two serialized methods (Torch~\cite{paszke2019pytorchimperativestylehighperformance} and DeepEP~\cite{deepep2025}). Specifically, Comet and UniEP explicitly split execution into three kernels: Dispatch+GroupGEMM, Activation, and GroupGEMM+Combine. FlashMoE and MegaMoE utilize a single fused kernel while maintaining a strict ordering between the two overlap stages. Conversely, Torch and DeepEP execute computation and communication serially with $G_{\mathrm{comm}} = G_{\mathrm{comp}} = 1$, representing a rigid, coarse-grained boundary between computation and communication.

\begin{table}[tbp]
    \centering
    \caption{Instantiation of the granularity abstraction for MoE EP strategies. Tile counts ($W_1$, $W_2$) and SM allocations ($N^{\mathrm{P}}_{\mathrm{SM}}$, $N^{\mathrm{M}}_{\mathrm{SM}}$) are strategy-specific.}
    \label{tab:granularity}
    \footnotesize
    \setlength{\tabcolsep}{2pt}
    \renewcommand{\arraystretch}{1.2}
    \begin{tabular}{@{}lcc@{}}
        \toprule
        \multicolumn{3}{c}{\textbf{Dispatch + GroupGEMM}} \\
        \midrule
        \textbf{Strategy}
        & $G_{\mathrm{comm}}$
        & $G_{\mathrm{comp}}$ \\
        \midrule
        Comet
        & $D-1$
        & $\min(\lceil\rho S_E/B_M\rceil,\lceil W_1/N_{\mathrm{SM}}\rceil)$ \\
        FlashMoE
        & $\lceil D E_{\mathrm{L}}/N^{\mathrm{M}}_{\mathrm{SM}}\rceil$
        & $\lceil W_1/N^{\mathrm{P}}_{\mathrm{SM}}\rceil$ \\
        MegaMoE
        & $\lceil S_{R_{\mathrm{in}}}k/N_{\mathrm{SM}}\rceil$
        & $\min(E_{\mathrm{L}}\lceil\rho S_E/B_M\rceil,\lceil W_1/N_{\mathrm{SM}}\rceil)$ \\
        UniEP
        & $\lceil S_{R_{\mathrm{in}}}k/N^{\mathrm{M}}_{\mathrm{SM}}\rceil$
        & $\min(E_{\mathrm{L}}\lceil\rho S_E/B_M\rceil,\lceil W_1/N^{\mathrm{P}}_{\mathrm{SM}}\rceil)$ \\
        Torch/DeepEP
        & $1$ & $1$ \\
        \midrule
        \multicolumn{3}{c}{\textbf{GroupGEMM + Combine}} \\
        \midrule
        \textbf{Strategy}
        & $G_{\mathrm{comp}}$
        & $G_{\mathrm{comm}}$ \\
        \midrule
        Comet
        & $\lceil W_2/N^{\mathrm{P}}_{\mathrm{SM}}\rceil$
        & $N_{\mathrm{split}}$ \\
        FlashMoE
        & $\lceil W_2/(N_{\mathrm{SM}}-1)\rceil$
        & $\lceil W_2/(N_{\mathrm{SM}}-1)\rceil$ \\
        MegaMoE
        & $\lceil W_2/N_{\mathrm{SM}}\rceil$
        & $\lceil W_2/N_{\mathrm{SM}}\rceil$ \\
        UniEP
        & $\lceil W_2/N^{\mathrm{P}}_{\mathrm{SM}}\rceil$
        & $\lceil S^{\max}_{R_{\mathrm{a}}}/N^{\mathrm{M}}_{\mathrm{SM}}\rceil$ \\
        Torch/DeepEP
        & $1$ & $1$ \\
        \bottomrule
    \end{tabular}
\end{table}

We examine the granularity of the four overlapping strategies. Let $N^{\mathrm{P}}_{\mathrm{SM}}$ and $N^{\mathrm{M}}_{\mathrm{SM}}$ denote the numbers of SMs allocated to computation and communication, respectively.
In Dispatch+GroupGEMM, Comet runs ring All-Gather on the CE, transferring one peer's full shard per step, so $G_{\mathrm{comm}}=D-1$. 
FlashMoE distributes $D E_{\mathrm{L}}$ transfer tasks, one per (peer, local-expert) pair, across $N^{\mathrm{M}}_{\mathrm{SM}}$ dispatch-specialized blocks, giving $G_{\mathrm{comm}}=\lceil D E_{\mathrm{L}}/N^{\mathrm{M}}_{\mathrm{SM}}\rceil$.
MegaMoE and UniEP both use $S_{R_{\mathrm{in}}}k$ token-level communication tasks. MegaMoE uses communication warps on every SM through warp specialization, with $G_{\mathrm{comm}}=\lceil S_{R_{\mathrm{in}}}k/N_{\mathrm{SM}}\rceil$, whereas UniEP uses dedicated communication SMs, with $G_{\mathrm{comm}}=\lceil S_{R_{\mathrm{in}}}k/N^{\mathrm{M}}_{\mathrm{SM}}\rceil$.
The first GroupGEMM processes $W_1=E_{\mathrm{L}}\lceil\rho S_E/B_M\rceil\lceil mI/B_N^{(1)}\rceil$ output tiles on the hottest rank. Computation requires $\lceil W_1/N_{\mathrm{SM}}\rceil$ waves, with $N_{\mathrm{SM}}$ replaced by $N^{\mathrm{P}}_{\mathrm{SM}}$ for FlashMoE and UniEP.
However, wave counts alone can overstate overlap: a tile becomes computable only after $B_M$ tokens of its expert arrive.
Comet receives tokens for all experts at the same rate, so a single hot expert bounds the rounds by $\lceil\rho S_E/B_M\rceil$, yielding $G_{\mathrm{comp}}=\min\!\left(\lceil\rho S_E/B_M\rceil,\lceil W_1/N_{\mathrm{SM}}\rceil\right)$.
MegaMoE and UniEP accumulate tokens expert by expert, relaxing this bound to $E_{\mathrm{L}}\lceil\rho S_E/B_M\rceil$. Thus, MegaMoE has
$G_{\mathrm{comp}}=\min(E_{\mathrm{L}}\lceil\rho S_E/B_M\rceil,$
$\lceil W_1/N_{\mathrm{SM}}\rceil)$, while UniEP has
$G_{\mathrm{comp}}=\min(E_{\mathrm{L}}\lceil\rho S_E/B_M\rceil,$
$\lceil W_1/N^{\mathrm{P}}_{\mathrm{SM}}\rceil)$.
FlashMoE's (peer, expert) task unit imposes no additional gating bound, so $G_{\mathrm{comp}}=\lceil W_1/N^{\mathrm{P}}_{\mathrm{SM}}\rceil$.

In GroupGEMM+Combine, the second GroupGEMM processes $W_2=E_{\mathrm{L}}\lceil\rho S_E/B_M\rceil\lceil H/B_N^{(2)}\rceil$ output tiles on the hottest rank. Both tile counts ($W_1$, $W_2$) and SM allocations ($N^{\mathrm{P}}_{\mathrm{SM}}$, $N^{\mathrm{M}}_{\mathrm{SM}}$) are strategy-specific.
Comet and UniEP use $N^{\mathrm{P}}_{\mathrm{SM}}$ computation SMs, giving $G_{\mathrm{comp}}=\lceil W_2/N^{\mathrm{P}}_{\mathrm{SM}}\rceil$.
For MegaMoE, $G_{\mathrm{comp}}=\lceil W_2/N_{\mathrm{SM}}\rceil$, as computation uses all SMs.
FlashMoE reserves one block for an in-kernel scheduler that never handles data, so $G_{\mathrm{comp}}=\lceil W_2/(N_{\mathrm{SM}}-1)\rceil$.
Communication differs most sharply.
Comet splits Combine along the output dimension into $N_{\mathrm{split}}$ rounds, with $G_{\mathrm{comm}}=N_{\mathrm{split}}$.
MegaMoE and FlashMoE fuse the return into the GEMM epilogue: each block directly pushes its completed tile, so communication is tile-granular and shares the computation round count.
Thus, $G_{\mathrm{comm}}=G_{\mathrm{comp}}=\lceil W_2/N_{\mathrm{SM}}\rceil$ for MegaMoE and $G_{\mathrm{comm}}=G_{\mathrm{comp}}=\lceil W_2/(N_{\mathrm{SM}}-1)\rceil$ for FlashMoE. 
UniEP instead returns each routed-token output through its communication SMs, yielding $G_{\mathrm{comm}}=\lceil S^{\max}_{R_{\mathrm{a}}}/N^{\mathrm{M}}_{\mathrm{SM}}\rceil$.

\subsection{Analytical Model Refinement}
\label{sec:analytical-refinement}

Eq.~\eqref{eq:overlap-time} can be inaccurate when neither granularity divides the other: misaligned chunk boundaries may delay computation until the next communication round completes. In practice, we therefore compute stage time round by round. For Dispatch+GroupGEMM, starting with $T=0$, we update
\begin{equation}
T \leftarrow
\max\!\left(
T,
\left\lceil \frac{jG_{\mathrm{comm}}}{G_{\mathrm{comp}}} \right\rceil
\frac{T_{\mathrm{comm}}}{G_{\mathrm{comm}}}
\right)
+\frac{T_{\mathrm{comp}}}{G_{\mathrm{comp}}},
\end{equation}
for $j=1,\ldots,G_{\mathrm{comp}}$. Each computation round starts only after the previous round finishes and all required data arrives. The final $T$ is the stage time. For GroupGEMM+Combine, we swap $\mathrm{comp}$ and $\mathrm{comm}$.

\section{Evaluation}
\subsection{Experiment Setup}

\subsubsection{Hardware Platforms}
The experiments were conducted on four NVIDIA GPU hardware platforms, covering three generations of mainstream data center GPU architectures: Ampere, Hopper, and Blackwell. Table~\ref{tab:Hardware} lists the key parameters of each platform.

\begin{table}[bp]
\vspace{-4mm}
\centering
\caption{Hardware platform specifications.}
\vspace{-3mm}
\label{tab:Hardware}
\footnotesize
\setlength{\tabcolsep}{4.5pt}
\begin{tabular}{@{}lcccccc@{}}
\toprule
\multirow{2}{*}{\textbf{Platform}}
  & \multicolumn{2}{c}{$F$ (TFLOPs)}
  & $B_{\mathrm{HBM}}$ & $B_{\mathrm{NVL}}$
  & \multirow{2}{*}{$N_{\mathrm{SM}}$}
  & \multirow{2}{*}{$D$} \\
\cmidrule(lr){2-3}
  & BF16 & FP8 & (GB/s) & (GB/s) & & \\
\midrule
A100  & 312  & --   & 2039 & 600  & 108 & 2/4/8 \\
H20   & 148  & 296  & 4022 & 900  & 78  & 2/4/8 \\
H800  & 989  & 1979 & 3350 & 400  & 132 & 2/4/8 \\
GB200 & 2500 & 5000 & 8000 & 1800 & 152 & 2/4   \\
\bottomrule
\end{tabular}
\vspace{-4mm}
\end{table}

\subsubsection{Kernel-Level Configurations}

We extracted and de-duplicated $(H,I,E,k)$ from the \texttt{config.json} files of all Hugging Face MoE checkpoints with at least $3{,}000$ downloads as of May 1, 2026, yielding $103$ shapes spanning $H\in[1024,8192]$, $I\in[256,5120]$, $E\in[8,512]$, and $k\in[2,12]$. These cover 7 LLM families, including DeepSeek-V3/V4~\cite{deepseekai2025deepseekv3technicalreport,deepseekai2026deepseekv4highlyefficientmilliontoken}, Qwen3~\cite{yang2025qwen3technicalreport}, GLM-4.x/5~\cite{glm5team2026glm5vibecodingagentic}, Kimi-K2~\cite{kimiteam2026kimik2openagentic}, Mixtral~\cite{jiang2024mixtralexperts}, gpt-oss~\cite{openai2025gptoss120bgptoss20bmodel}, and Phi-3~\cite{abdin2024phi3technicalreporthighly}.

We test each shape at $S_{R_{\mathrm{in}}}\in\{32,64,128,256,512,1\text{K},2\text{K},4\text{K},\\8\text{K},16\text{K}\}$, covering decode-scale batches through 8K-token chunked prefill. Of $1030\times20=20{,}600$ configuration--strategy--platform combinations, $13{,}734$ ($66.7\%$) are measurable. The rest violate implementation constraints (e.g.\ constraints on $E$, $k$ or $B_M$ divisibility). Feasibility is a property of the implementation, not a choice on our part: a strategy is never penalized on a configuration it cannot run.

Differences in supported precision, MLP structure, and platform require two experiment groups (Table~\ref{tab:Strategy}). Before timing, we validate every strategy on all measured configurations against two serial methods: Torch (Group~I) or DeepEP (Group~II). Relative $\ell_2$ errors remain below $10^{-2}$ for BF16 and $5\times10^{-2}$ for FP8, consistent with accumulation-order and quantization effects. All speedups thus compare functionally equivalent implementations.

\begin{table}[bp]
\centering
\caption{Two experiment groups to evaluate OverlapAdvisor.}
\vspace{-3mm}
\label{tab:Strategy}
\footnotesize
\setlength{\tabcolsep}{4pt}

\renewcommand{\arraystretch}{1.15}
\begin{tabularx}{\columnwidth}{@{}l
  >{\raggedright\arraybackslash}X
  >{\raggedright\arraybackslash}X@{}}
\toprule
 & \textbf{Group~I} & \textbf{Group~II} \\
\midrule
Precision  & BF16 & FP8 \\
MLP        & Standard, GELU & Gated, SwiGLU \\
Routing    & Non-uniform, $\rho=2.0$ & Non-uniform, $\rho=2.0$ \\
Timing     & CUDA Graph\textsuperscript{a} & CUDA Graph\textsuperscript{a} \\
\midrule
A100       & Comet, FlashMoE, Torch & -- \\
H20, H800  & Comet, FlashMoE, Torch, DeepEP-v1\textsuperscript{b}
           & MegaMoE\textsuperscript{c}, UniEP, DeepEP-v1\textsuperscript{b} \\
GB200      & -- & MegaMoE, UniEP, DeepEP-v2 \\
\bottomrule
\end{tabularx}
\vspace{2pt}
{\footnotesize\raggedright
\textsuperscript{a}Comet and DeepEP-v1 Normal are timed with CUDA Event.\par
\textsuperscript{b}DeepEP-v1 uses LowLatency (LL) for $S\le256$ and Normal otherwise, fitted separately but treated as one strategy by the selector.\par
\textsuperscript{c}Cross-architecture ports: MegaMoE on Hopper follows the SM90 megakernel design of~\cite{MegaMoE90}, and UniEP on 
Blackwell is ported from Hopper. Both preserve the original overlap structure.\par}
\end{table}

\begin{figure}[!htbp] 
\centering 
\includegraphics[width=0.48\textwidth]{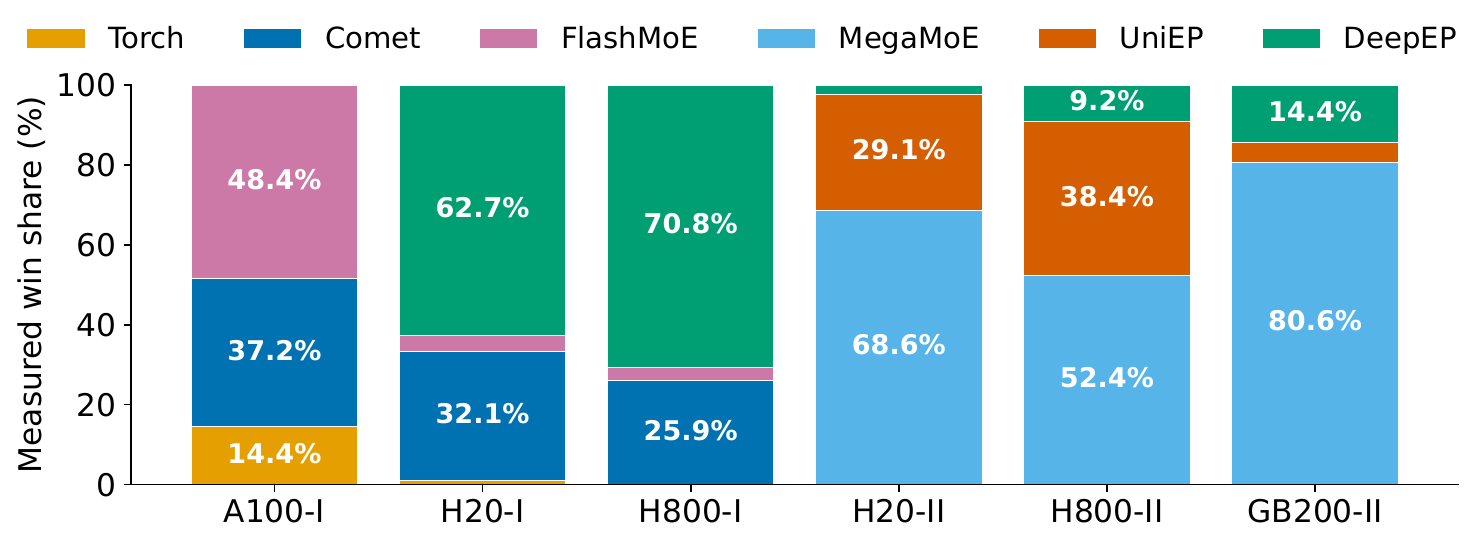}
\vspace{-8mm}
\caption{Performance win rates of MoE EP strategies across test configurations.} 
\label{fig:implementation} 
\vspace{-4mm}
\end{figure}

Figure~\ref{fig:implementation} illustrates the proportion of test configurations in which each strategy performs best, highlighting distinct regions of advantage. Across all platforms, no single strategy achieves optimality in more than 80.6\% of cases.

\subsubsection{Data-Driven Model Configurations}
\label{sec:Data-Driven-Config}

We fit each strategy on a random $20\%$ of measured samples (averaging $137$ per strategy--hardware pair) by minimizing MAPE with DE, and evaluate on the remaining $80\%$. Five seeds control both data splitting and DE initialization. This small training fraction is deliberate: the $17$--$23$ physically interpretable parameters per strategy (bandwidth utilization and compute efficiency) are far fewer than the training samples, limiting overfitting risk, while the large remaining set tests generalization to unseen configurations.

Table~\ref{tab:DE} lists kernel decompositions and DE budgets. For  kernels with $\le4$, $5$--$9$, and $\ge10$ parameters, we empirically set $(N_p,\mathrm{maxiter})$ to $(15,80)$, $(20,120)$, and $(30,200)$, respectively, where $N_p$ is population size and $\mathrm{maxiter}$ is the iteration limit. All use mutation-factor dithering and $CR=0.9$. Fitting all $90$ instances across four GPU platforms takes about $13$ minutes on an 8-vCPU Intel Xeon host (2.50 GHz, 16 GB memory), negligible compared with measurement collection.

\begin{table*}[!htbp]
\centering
\caption{Kernel decomposition and DE budget of each strategy. $|\theta|_{\mathrm{k}}$ denotes the number of fitted parameters per kernel, and $|\theta|_{\mathrm{tot}}$ denotes the total number per strategy.}
\vspace{-3mm}
\label{tab:DE}
\footnotesize
\setlength{\tabcolsep}{4pt}
\renewcommand{\arraystretch}{1.05}
\begin{tabular}{@{}llcclc@{}}
\toprule
\textbf{Strategy}
& \textbf{Partitioned Kernels}
& \textbf{Kernels}
& $\bm{|\theta|_{\mathrm{k}}}$
& $\bm{(N_p, \mathrm{maxiter})}$
& $\bm{|\theta|_{\mathrm{tot}}}$ \\
\midrule
Comet / UniEP
& Dispatch+GroupGEMM / Act / GroupGEMM+Combine
& 3 & 3--7
& (15,\,80), (20,\,120)
& 17 \\

Torch
& AG / permute / GroupGEMM / Act / GroupGEMM+unpermute+RS
& 5 & 3--10
& (15,\,80), (30,\,200)
& 23 \\

FlashMoE / MegaMoE
& MegaKernel
& 1 & 17
& (30,\,200)
& 17 \\

DeepEP-v1 Normal
& Dispatch / permute / GroupGEMM / Act / GroupGEMM / unpermute / Combine
& 7 & 3--4
& (15,\,80)
& 23 \\

DeepEP-v1 LL
& Dispatch / GroupGEMM / Act / GroupGEMM / Combine
& 5 & 3--4
& (15,\,80)
& 17 \\

DeepEP-v2
& Dispatch / GroupGEMM+Act+GroupGEMM / Combine
& 3 & 3--11
& (15,\,80), (30,\,200)
& 17 \\
\bottomrule
\end{tabular}
\end{table*}

\subsubsection{E2E-Level Configurations}
\label{sec:E2E-level-Configurations}

We evaluate Group~II (Table~\ref{tab:Strategy}) on one fully NVLink-connected $8\times$H20 node using SGLang v0.5.15-post1 and vLLM~0.28.0, with CUDA~13.0 (nvcc~12.8), PyTorch~2.13.0, and NCCL~2.30.7. Both use DP attention ($\mathrm{dp\_size}=D$), $\mathrm{EP}=D$, and $\mathrm{TP}=1$. (1) SGLang uses FlashInfer attention, Mooncake PD KV transfer, and a PD-aware router; (2) vLLM uses FlashAttention-3/FA-MLA, NIXL over NVLink, and an OpenAI-compatible proxy. 

The three Group~II strategies replace only the MoE layer and are selected via launch flags or environment variables. BF16 checkpoints are quantized to block-FP8 at load time; Qwen3-235B-A22B-FP8 uses native block-FP8. Each scenario starts a fresh server with overlap/async scheduling, prefix caching, and shared-expert fusion disabled; decode CUDA graphs enabled; an $8192$-token per-rank prefill budget; and a GPU memory budget of $80\%$ per device.

Both engines share the models and deployments in Table~\ref{tab:e2e}. SGLang replays the Mooncake production trace~\cite{qin2025mooncake}, and vLLM replays ShareGPT~V3~\cite{sharegpt-v3} conversations, preserving natural length distributions under PD disaggregation. Colocated deployments use two subsets per trace: prefill-dominated and decode-dominated. We run $32D$ concurrent requests in total ($32$ per rank) and report the median of three runs per scenario after one warm-up.

\begin{table}[!htbp]
  \centering
  \caption{Models and deployments used in E2E evaluation. PD $a{+}b$ denotes $a$ GPUs for prefill and $b$ GPUs for decode.}
  \vspace{-3mm}\label{tab:e2e}
  \footnotesize
  \setlength{\tabcolsep}{3pt}
  \renewcommand{\arraystretch}{1.05}
  \begin{tabular}{@{}clcc@{}}
  \toprule
  \textbf{ID} & \textbf{Model} & $\bm{H/I/E/k}$ & \textbf{EP degree} $\bm{(D)}$ \\
  \midrule
  E2E-1 & Qwen3-30B-A3B  & 2048/768/128/8   & PD $2{+}2$ (2) \\
  E2E-2 & GLM-4.7-Flash  & 2048/1536/64/4   & PD $2{+}2$ (2) \\
  E2E-3 & Qwen3-Next-80B & 2048/512/512/10  & PD $4{+}4$ (4) \\
  \midrule
  E2E-4 & \multirow{2}{*}{Qwen3-235B-A22B-FP8}
        & \multirow{2}{*}{4096/1536/128/8}
        & \multirow{2}{*}{colocated (8)} \\
  E2E-5 & & & \\
  \bottomrule
  \end{tabular}
  \vspace{-3mm}
  \end{table}

\subsubsection{Baselines}
We compare OverlapAdvisor with six fixed-strategy baselines (Table~\ref{tab:Strategy}) at the kernel level and three of them (MegaMoE, UniEP, and DeepEP) end to end. Each baseline uses one strategy across configurations and, under PD disaggregation, for both prefill and decode.

\subsubsection{Metrics}
\leavevmode\par

\textbf{Performance Improvement.}
Speedup over each baseline is $T_{\mathrm{baseline}}/T_{\mathrm{OverlapAdvisor}}$ for kernels and $Q_{\mathrm{OverlapAdvisor}}/Q_{\mathrm{baseline}}$ end to end, where $T$ is execution time ($\mu$s) and $Q$ is throughput (tokens/s); values above $1$ indicate improvement. Kernel speedups are averaged over test configurations per seed; we report the mean and min--max range across five seeds.

\textbf{Execution Time Regret.}
For each test configuration,
\begin{equation}
R_i = \left(\frac{T_i}{T_i^{\mathrm{best}}}-1\right)\times 100\%,
\end{equation}
where $T_i^{\mathrm{best}}$ is the fastest measured kernel execution time among feasible strategies. A CDF closer to the upper-left indicates lower regret across more configurations.

\textbf{Prediction Error.}
We use mean absolute percentage error:
\begin{equation}
\mathrm{MAPE} = \frac{100\%}{n}
\sum_{i=1}^{n}\frac{|\widehat{T}_i-T_i|}{T_i},
\end{equation}
where $\widehat{T}_i$ and $T_i$ are the predicted and measured execution times for test configuration $i$, and $n$ is the test-set size per strategy--hardware pair and seed. We report mean MAPE and its min--max range across five seeds for each pair.

\textbf{Selection Accuracy.}
Top-1 accuracy is the fraction of test configurations where OverlapAdvisor selects the fastest strategy, reported as the mean and min--max range across five seeds at each token batch size $S_{R_{\mathrm{in}}}$.

\subsection{Results and Analysis}
\label{sec:Evaluation}

\subsubsection{Kernel-Level Performance}
\label{sec:Kernel-level-Performance}
\leavevmode\par

\begin{figure}[bp]
\vspace{-3mm}
\centering
\includegraphics[width=0.48\textwidth]{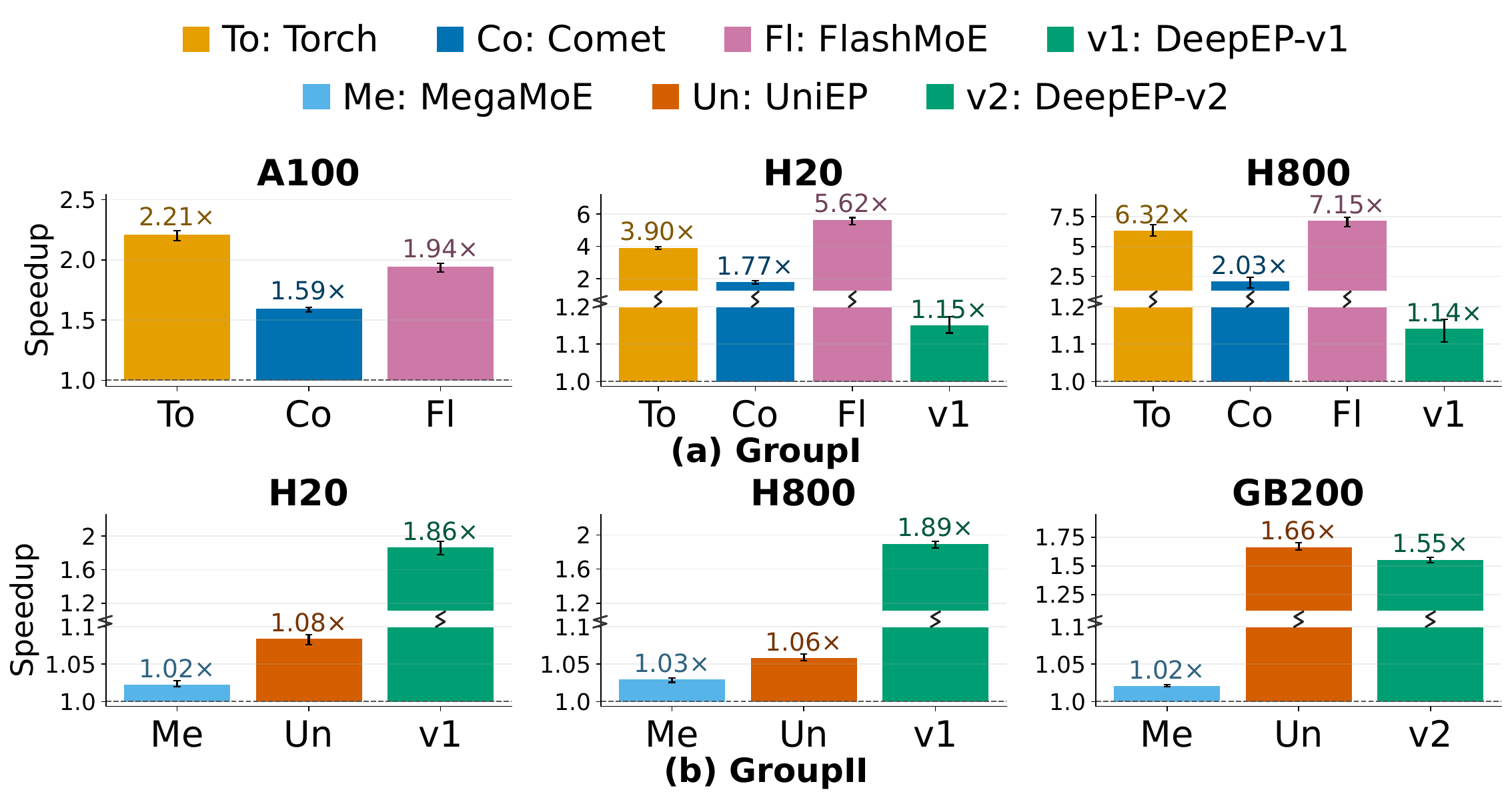}
\vspace{-8mm}
\caption{OverlapAdvisor kernel-level speedup over baselines.}
\label{fig:evaluation_kernel1}
\end{figure}

\textbf{Kernel Execution Time.}
Figures~\ref{fig:evaluation_kernel1}(a)--(b) compare OverlapAdvisor with six fixed-strategy baselines in execution-time speedup. (1) OverlapAdvisor outperforms all available fixed strategies, achieving $5.62\times$ over the FlashMoE on GroupI-H800 and $1.86\times$ over the DeepEP-v1 on GroupII-H20.
(2) Against the strongest fixed strategy on each platform, OverlapAdvisor achieves $1.59\times$ speedup over Comet on GroupI-A100; $1.15\times$ and $1.14\times$ over DeepEP-v1 on GroupI-H20 and GroupI-H800; and $1.02\times$--$1.03\times$ over MegaMoE on the three Group~II platforms. 
(3) Across both experiment groups, the average gain is $1.16\times$, peaking at $1.59\times$.

\begin{figure}[bp]
\vspace{-3mm}
\centering
\includegraphics[width=0.48\textwidth]{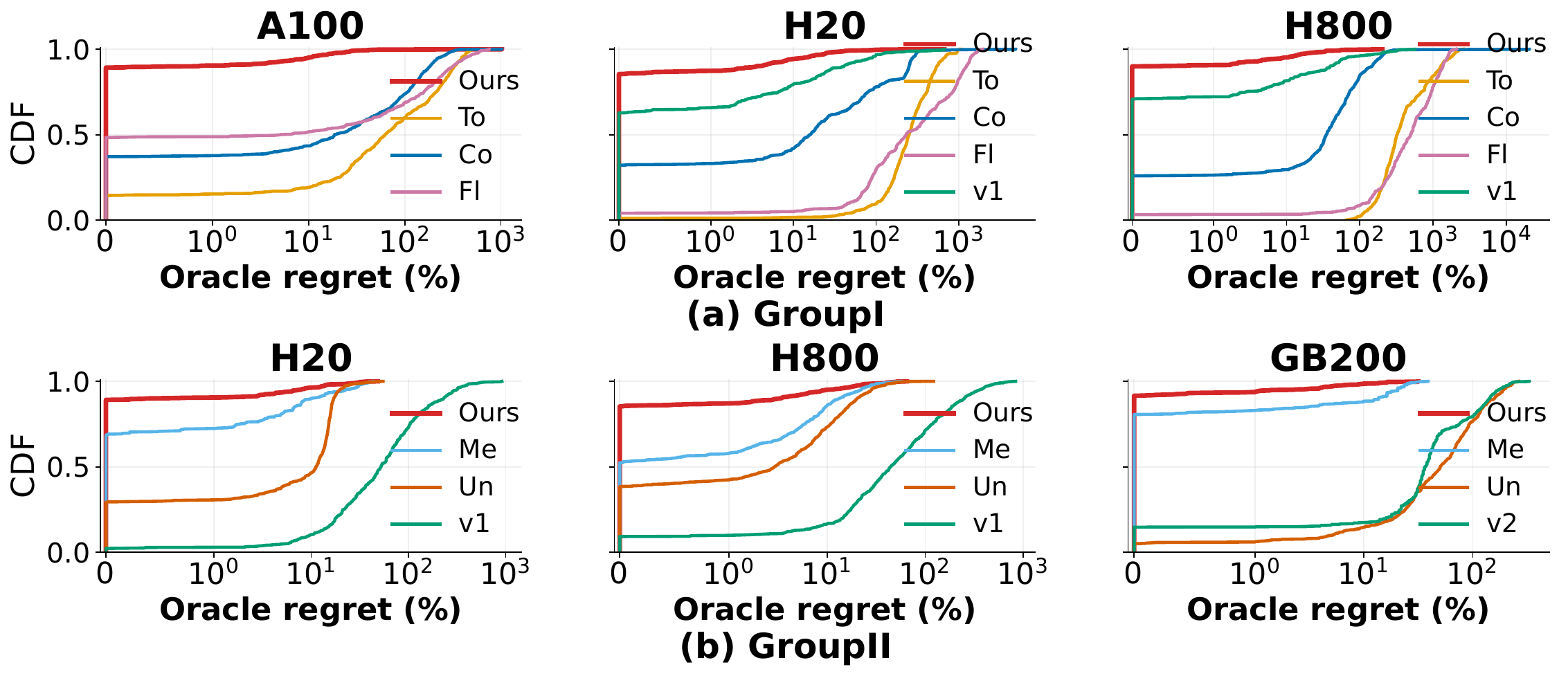}
\vspace{-5mm}
\caption{Kernel regret CDF of OverlapAdvisor and baselines.}
\label{fig:evaluation_kernel2}
\end{figure}

The modest $2$--$3\%$ average gains over MegaMoE on Group~II reflect many configurations where MegaMoE is already fastest and selection offers no gain. However, OverlapAdvisor outperforms MegaMoE by more than $5\%$ on 107/773 ($13.84\%$), 184/772 ($23.83\%$), and 55/504 ($10.91\%$) test configurations on GroupII-H20, GroupII-H800, and GroupII-GB200, respectively (Fig.~\ref{fig:evaluation_kernel2}). Thus, CSA avoids substantial performance losses in configurations where MegaMoE is suboptimal.

\subsubsection{Across-Engine E2E-Level Performance}
\label{sec:E2E-level-Performance}

\begin{figure}[htbp] 
\vspace{-2mm}
\centering 
\includegraphics[width=0.48\textwidth]{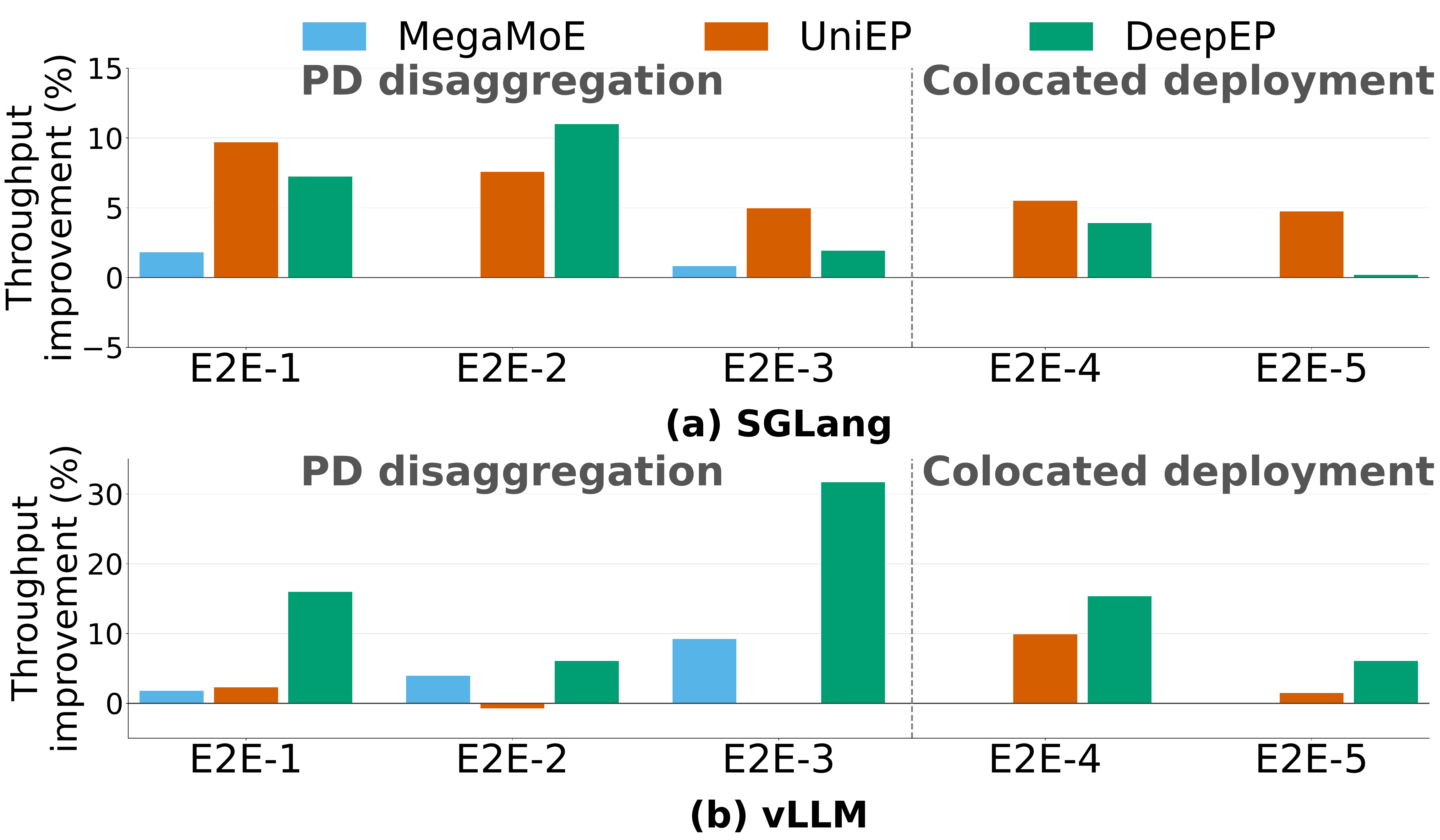}
\vspace{-7mm}
\caption{E2E-Level Evaluation: OverlapAdvisor throughput improvement over baselines (\%).} 
\label{fig:evaluation_e2e} 
\vspace{-4mm}
\end{figure}

Across five deployments on two serving engines (Figure~\ref{fig:evaluation_e2e}), OverlapAdvisor improves throughput over fixed-strategy baselines MegaMoE, UniEP, and DeepEP by averages of $1.78\%$, $4.57\%$, and $9.97\%$, respectively, with maxima of $9.28\%$, $9.95\%$, and $31.72\%$. These results demonstrate the necessity of strategy selection across different configurations and serving engines.

\subsection{Overhead}

OverlapAdvisor incurs a one-time fitting cost of approximately $13$ minutes per seed across all evaluated strategies and platforms on an 8-vCPU Intel Xeon host. At server launch, strategy selection takes approximately $6$~$\mu$s without online profiling or warm-up. The performance-model code and fitted parameters for all $24$ models together occupy approximately 26 KB of storage. The selected backend remains fixed during serving, requiring no profiling or runtime selection.

\subsection{Ablation Study}

\subsubsection{Prediction Accuracy of UPP} 
\begin{figure}[tbp]
\centering
\includegraphics[width=0.48\textwidth]{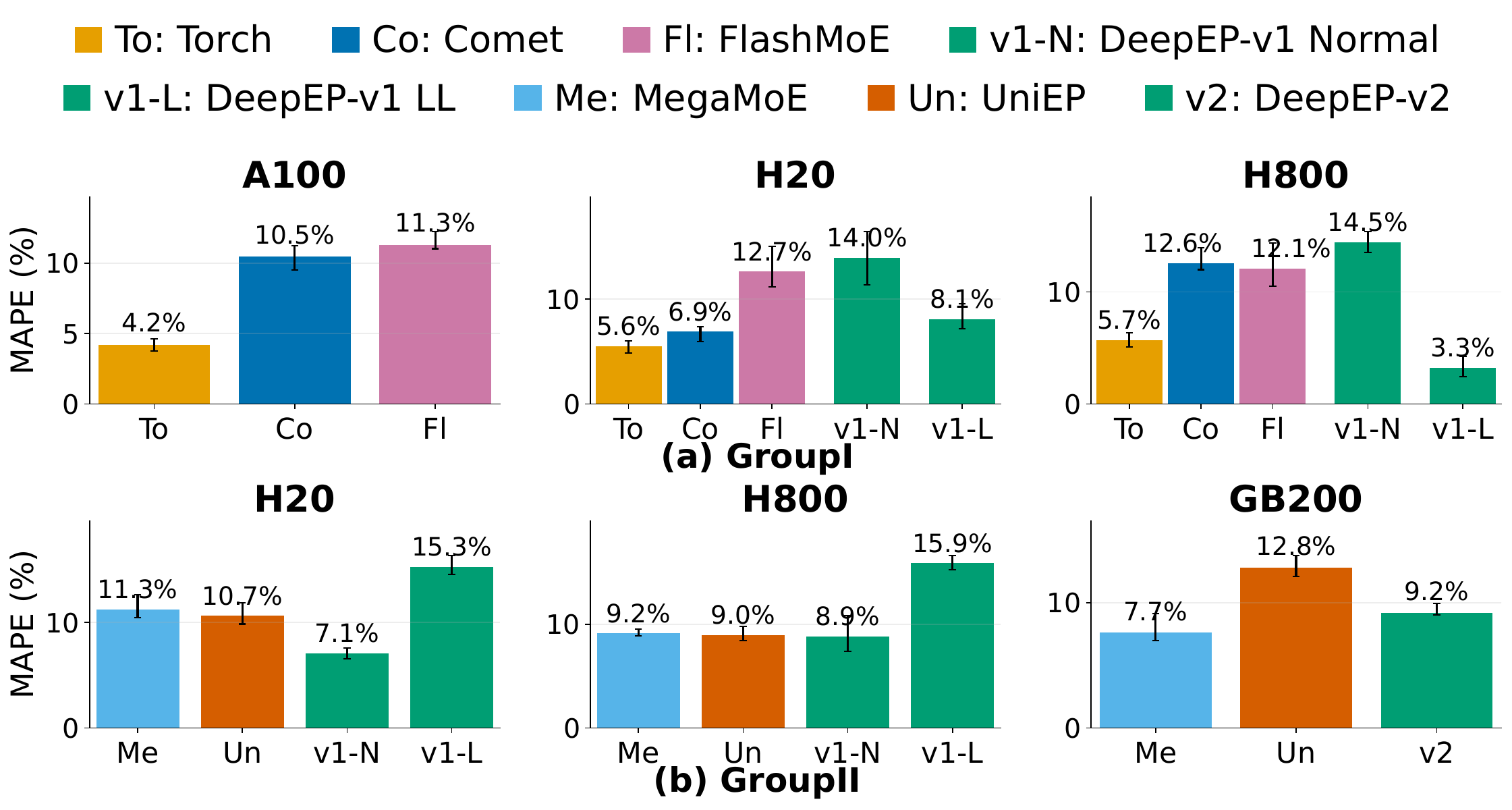}
\vspace{-7mm}
\caption{Ablation Study: Prediction MAPE of UPP.}
\label{fig:evaluation_ablation1}
\vspace{-4mm}
\end{figure}

Figure~\ref{fig:evaluation_ablation1} reports UPP's MAPE across 20 $\langle$strategy, platform$\rangle$ pairs. DeepEP-v1's LL and Normal modes are fitted separately but treated as one strategy by CSA, yielding 24 models across 8 implementations of 6 strategies. Errors range from $3.3\%$ to $15.9\%$ (median $9.9\%$), demonstrating UPP's effectiveness in modeling complex kernels such as FlashMoE and MegaMoE.

\subsubsection{Predictor Comparison and Component Ablation of UPP}
We compare UPP with learned predictors and ablate its analytical and data-driven models (Table~\ref{tab:ablation}), reporting mean MAPE across $\langle$strategy, platform$\rangle$ pairs under two data splits. The random split follows \S\ref{sec:Data-Driven-Config}, training on $20\%$ of the configurations. The extrapolation split trains on the four smallest token batches ($S_{R_{\mathrm{in}}}\le256$, $40\%$ of configurations) and tests on the six largest ($S_{R_{\mathrm{in}}}\ge512$). All fitted predictors use identical training and test data within each split. Fixed-efficiency variants use $\eta_{\mathrm{gemm}}=\eta_{\mathrm{mem}}=\eta_{\mathrm{comm}}=0.80$ and are evaluated on the same random-split test set.

\begin{table}[bp]
\vspace{-2mm}
\centering
\caption{Prediction MAPE (\%, lower is better).
(a) OverlapAdvisor versus learned predictors, all fitted.
(b) Ablation of analytical model and data-driven model.}
\label{tab:ablation}
\vspace{-3mm}
\small
\setlength{\tabcolsep}{4pt}

\textbf{(a) Model effectiveness}\par
\vspace{3pt}
\begin{tabular*}{\columnwidth}
{@{\extracolsep{\fill}}lcc@{}}
\toprule
Predictor & Random split & Extrapolation \\
\midrule
OverlapAdvisor & \textbf{9.76} & \textbf{30.97} \\
XGBoost    & 13.23         & 53.85 \\
MLP        & 13.30         & 49.90 \\
\bottomrule
\end{tabular*}
\par\vspace{6pt}

\textbf{(b) Analytical and data-driven model effectiveness}\par
\vspace{3pt}
\begin{tabular*}{\columnwidth}
{@{\extracolsep{\fill}}lccc@{}}
\toprule
 & \multicolumn{2}{c}{Random split}
 & Extrapolation \\
\cmidrule(lr){2-3}\cmidrule(l){4-4}
\shortstack[l]{Analytical model}
 & Fixed & Fitted & Fitted \\
\midrule
Granularity-aware (OverlapAdvisor)
    & \textbf{55.68} & \textbf{9.76} & \textbf{30.97} \\
Serialized
    & 62.95 & 9.99 & 39.31 \\
Perfect overlap
    & 59.08 & 9.89 & 35.45 \\
\bottomrule
\end{tabular*}
\end{table}

\textbf{Model Effectiveness.}
XGBoost and MLP use configurations, hardware specifications, and derived quantities $V_{\mathrm{comm}}$, $V_{\mathrm{mem}}$, and $C_{\mathrm{gemm}}$, but not our granularity or efficiency functions. As Table~\ref{tab:ablation}(a) shows, OverlapAdvisor achieves $9.76\%$ MAPE on the random split, versus $13.23\%$ and $13.30\%$ for XGBoost and MLP. Its advantage grows under extrapolation: $30.97\%$ versus $53.85\%$ and $49.90\%$. With limited training data, these predictors must learn execution time relationships from samples, whereas OverlapAdvisor explicitly models how volumes, granularities, and efficiencies compose.

\textbf{Analytical Model Effectiveness.}
We replace the analytical model (Eq.~\eqref{eq:overlap-time}) with serialized execution or perfect overlap, leaving the remaining model unchanged (Table~\ref{tab:ablation}(b)). With fixed efficiencies, OverlapAdvisor achieves $55.68\%$ MAPE, below both alternatives ($62.95\%$ and $59.08\%$). After fitting, OverlapAdvisor retains an advantage of up to $0.23$ percentage points on the random split. Although fitted efficiency functions partially compensate for analytical model errors, this compensation lacks a physical basis and generalizes poorly under extrapolation: OverlapAdvisor achieves $30.97\%$ MAPE versus $39.31\%$ and $35.45\%$, an advantage of up to $8.34$ percentage points. Together, these three settings demonstrate the effectiveness of OverlapAdvisor's analytical overlap structure.

\textbf{Data-Driven Model Effectiveness.}
On the same random-split test set, fitting reduces OverlapAdvisor's MAPE from $55.68\%$ to $9.76\%$ (Table~\ref{tab:ablation}(b)). This demonstrates the importance of platform- and strategy-specific parameter fitting.

\subsubsection{Training Data Ratio of UPP}

\begin{figure}[!htbp]
\centering 
\includegraphics[width=0.48\textwidth]{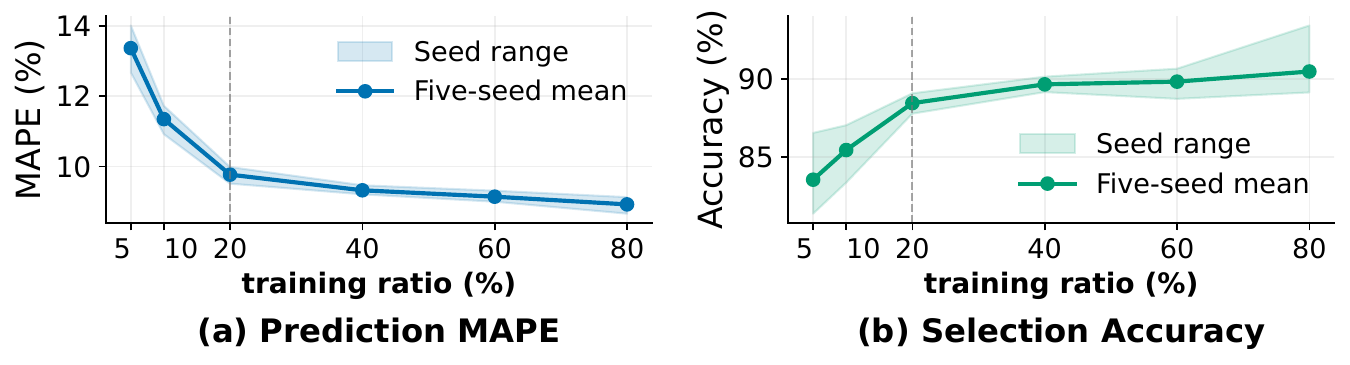}
\vspace{-8mm}
\caption{Training efficiency and accuracy of UPP.} 
\label{fig:evaluation_ablation2} 
\vspace{-1mm}
\end{figure}

Following the protocol in Section~\ref{sec:Data-Driven-Config}, we vary the training ratio over $5\%$, $10\%$, $20\%$, $40\%$, $60\%$, and $80\%$, reporting average MAPE and strategy-selection accuracy. Figure~\ref{fig:evaluation_ablation2} shows the mean and min--max range across five random seeds.

As the training ratio increases from $5\%$ to $20\%$, UPP's MAPE drops from $13.36\%$ to $9.76\%$, and CSA's accuracy rises from $83.52\%$ to $88.43\%$; both curves flatten thereafter. At $20\%$, each seed uses about $2.75$K hardware--strategy samples and takes $13$ minutes on an eight-core Intel Xeon 2.50\,GHz CPU. Raising the ratio to $40\%$ doubles profiling to $5.50$K samples for marginal gains. We therefore select $20\%$ using the elbow method to balance prediction accuracy, selection accuracy, and profiling overhead. OverlapAdvisor generalizes from this limited train set to the remaining $80\%$ of configurations, supporting the effectiveness of UPP and CSA.

\subsubsection{Selection Accuracy of CSA}
\begin{figure}[tbp]
\centering
\includegraphics[width=0.47\textwidth]{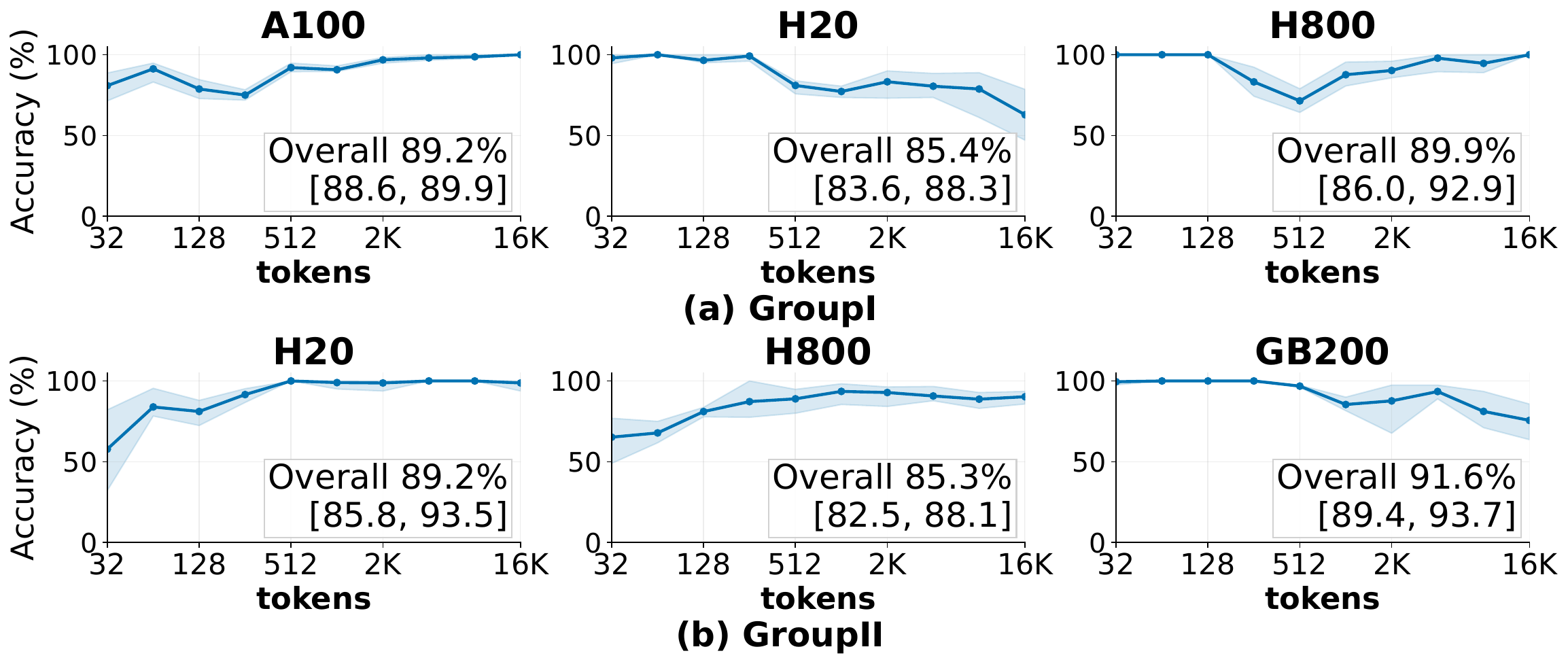}
\vspace{-3mm}
\caption{Ablation Study: Selection Accuracy of CSA.}
\label{fig:evaluation_ablation3}
\vspace{-4mm}
\end{figure}

Figure~\ref{fig:evaluation_ablation3} shows top-1 selection accuracy versus token size. Accuracies are $89.2\%$, $85.4\%$, and $89.9\%$ for GroupI-A100/H20/H800, and $89.2\%$, $85.3\%$, and $91.6\%$ for GroupII-H20/H800/GB200, respectively, giving an average accuracy of $88.43\%$ across both experiment groups.

Accuracy is generally lower when leading strategies have nearly identical performance, such as MegaMoE and UniEP at small token size on GroupII-H20 and GroupII-H800, or MegaMoE and DeepEP-v2 at large token size on GroupII-GB200. In these near-tie regimes, selecting the second-fastest strategy incurs little penalty, so occasional top-1 errors have limited impact on the speedups and regret in Fig.~\ref{fig:evaluation_kernel1} and Fig.~\ref{fig:evaluation_kernel2}.

\subsection{Analysis and Optimization of Existing Strategies}

\begin{figure}[tbp] 
\centering 
\includegraphics[width=0.48\textwidth]{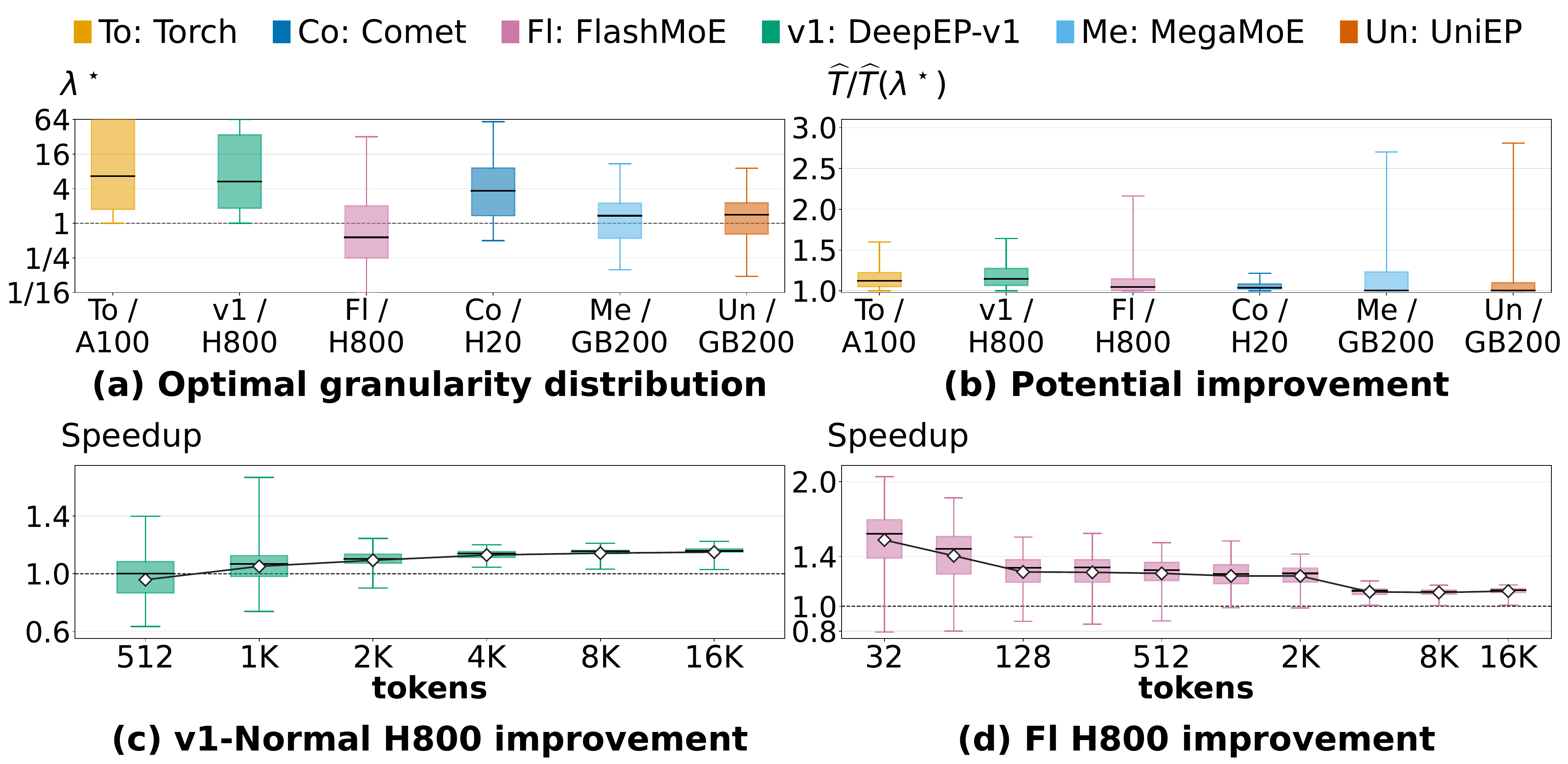}
\vspace{-6mm}
\caption{Optimal granularity distribution, potential improvement and realized improvement.} 
\label{fig:case} 
\vspace{-4mm}
\end{figure}

Beyond strategy selection, OverlapAdvisor can diagnose bottlenecks related to granularity by assessing whether native granularities effectively balance overlap and efficiency, while identifying opportunities for tuning. With fitted efficiency parameters fixed, we scale native granularities as $G_j(\lambda)=\max(1,\lambda G_j^{\mathrm{native}})$ and find the $\lambda^\star$ minimizing predicted execution time. Finer partitioning improves overlap but reduces per-round volume, potentially lowering communication and memory efficiency. Figures~\ref{fig:case}(a) and (b) show the distributions of $\lambda^\star$ and potential improvement relative to native granularities, respectively, revealing three regimes.

\textbf{Coarse granularity limits overlap.}
Torch and DeepEP-v1 lack native overlap, with median $\lambda^\star$ values of $6.58$ and $5.30$ and median potential improvement of $1.123 \times$ and $1.149 \times$, respectively. Comet supports overlap but remains too coarse: communication limits its two stages to $D-1$ ring steps and $N_{\mathrm{split}}$ combine rounds, preventing pipeline depth from scaling with workload. Refinement yields a median $\lambda^\star=3.67$, but only $1.036 \times$ median headroom.

Guided by this diagnosis, we refine DeepEP-v1 by partitioning tokens into $g$ groups and dispatching them in group-major order. We choose $g$ from $2$ to $8$ depending on the token count (e.g., $g=2$ for $512$ tokens and $g=8$ for $16K$ tokens) to balance computation efficiency and overlap. Per-group arrival counters allow each group's GroupGEMM to start as soon as its inputs arrive, overlapping computation with the remaining dispatch. This increases $G^{(1)}_{\mathrm{comp}}=G^{(1)}_{\mathrm{comm}}$ from $1$ to $g$. As shown in Figure~\ref{fig:case}(c), the refinement achieves a median speedup of $1.099\times$, reaching $1.158\times$ at $S_{R_{\mathrm{in}}}=16K$ and improving $511$ of $594$ configurations.

\textbf{Excessive granularity reduces efficiency.}
Only FlashMoE has a median $\lambda^\star<1$ ($0.57$), with a median predicted speedup of $1.047\times$ from coarsening. On H800-I with $(S_{R_{\mathrm{in}}},H,I,$ $E,k,D)=(4096,4096,2048,128,8,8)$, $\rho=2$, and $B_M=B_N=128$, Dispatch+GroupGEMM splits the $131$ non-scheduler SMs into $98$ compute and $33$ communication blocks, giving $G^{(1)}_{\mathrm{comp}}=\lceil W_1/98\rceil=84$ and $G^{(1)}_{\mathrm{comm}}=\lceil DE_{\mathrm{L}}/33\rceil=4$. In GroupGEMM + Combine, each block computes and returns one output tile, yielding $G^{(2)}_{\mathrm{comp}}=G^{(2)}_{\mathrm{comm}} =\lceil W_2/131\rceil=126$.

We coarsen only second-stage communication by grouping each row block's $\lceil H/B_N^{(2)}\rceil=32$ tiles under one signal. This increases the Combine payload by $32\times$, cuts signals by $32\times$, and reduces $G^{(2)}_{\mathrm{comm}}$ from $126$ to $\lceil E_{\mathrm{L}}\lceil\rho S_E/B_M\rceil/131\rceil=4$. Communication efficiency improves, but the overlap divisor $\min(G^{(2)}_{\mathrm{comp}},G^{(2)}_{\mathrm{comm}})$ in Eq.~\eqref{eq:overlap-time} drops from $126$ to $4$. OverlapAdvisor identifies when the efficiency gain outweighs this overlap loss. Across $1030$ measured configurations on $8\times$H800, coarsening achieves a median speedup of $1.217\times$, improving $910$ configurations (Figure~\ref{fig:case}(d)).

\textbf{Granularity is not always the bottleneck.}
MegaMoE and UniEP lie near flat optima: their median $\lambda^\star$ values are $1.35$ and $1.41$, but median potential improvements are only $1.017 \times$ and $1.015 \times$. On GB200-II with $(S_{R_{\mathrm{in}}},H,I,E,k,D)=(8\text{K},2048,512,512,10,4)$, MegaMoE takes $2.328$\,ms versus DeepEP-v2's $1.862$\,ms despite much finer granularities. Its effective (GEMM, HBM, NVLink) efficiencies are $(0.576,0.701,$ $0.598)$ versus $(0.763,0.790,0.804)$; granularity tuning alone cannot close this efficiency gap.

\section{Related Work}

We group related work into kernel modeling, overlap optimization, and strategy integration (Table~\ref{tab:related-work}).

\textbf{Kernel modeling.}
Roofline~\cite{10.1145/1498765.1498785} identifies compute and memory bottlenecks through arithmetic intensity and peak performance bounds. PipeWeave~\cite{zhang2026pipeweavesynergizinganalyticallearning} combines analytical pipeline-cycle estimates with a lightweight MLP for cross-pipeline coupling. Both focus on individual kernels rather than cross-strategy EP overlap.

\textbf{Overlap optimization.}
Lagom~\cite{xu2026lagomunleashingpowercommunication} uses a unified overlap loss to tune kernel communication parameters. Tutel~\cite{hwang2023tuteladaptivemixtureofexpertsscale} supports low-cost parallelism switching through an invariant expert-weight layout and tunes pipeline depth at the engine/runtime layer. Neither provides a unified performance model across diverse EP overlap implementations.

\textbf{Strategy integration.}
DynaFlow~\cite{pan2026dynaflowtransparentflexibleintradevice} decouples model definitions from execution schedules to integrate intra-device parallelism strategies, but provides no predictive selector.

OverlapAdvisor instead provides a unified EP overlap model for strategy selection and granularity diagnosis, validated across MoE workloads and hardware.

\begin{table}[tbp]
    \centering
    \caption{Comparison of related work.}
    \vspace{-3mm}
    \label{tab:related-work}
    \footnotesize
    \setlength{\tabcolsep}{3pt}
    \renewcommand{\arraystretch}{1.1}
    \begin{tabular}{@{}m{0.21\columnwidth}
                        m{0.17\columnwidth}
                        m{0.12\columnwidth}
                        m{0.40\columnwidth}@{}}
        \toprule
        \textbf{Work} & \textbf{Scope} & \textbf{Layer} & \textbf{Function} \\
        \midrule
        Roofline  & Kernel      & Kernel   & Performance model \\
        PipeWeave & Kernel      & Kernel   & Performance model \\
        Lagom     & Overlap     & Kernel   & Parameter tuning \\
        Tutel     & MoE overlap & Engine   & Mode switching \\
        DynaFlow  & Overlap     & Strategy & Strategy integration \\
        \midrule
        \textbf{OverlapAdvisor}
        & \textbf{EP overlap}
        & \textbf{Strategy}
        & \textbf{Performance model, strategy adaptation and tuning} \\
        \bottomrule
    \end{tabular}
    \vspace{-2mm}
\end{table}

\section{Conclusion and Future Work}

OverlapAdvisor unifies performance modeling for MoE EP strategies, enabling selection and optimization in two serving engines. First, it predicts performance with $3.3\%$--$15.9\%$ MAPE. Second, its launch-time selection takes about $6$~$\mu$s without profiling, achieving $88.43\%$ top-1 accuracy. Third, it delivers $1.16\times$ average and $1.59\times$ maximum kernel speedups over the best fixed strategies. Fourth, it improves end-to-end throughput over MegaMoE, UniEP, and DeepEP by averages of $1.78\%$, $4.57\%$, and $9.97\%$, with gains up to $31.72\%$.

Future work will extend OverlapAdvisor to multi-node deployments, model SM- and warp-level resource contention among overlapping kernels, and further unify prediction across platforms and strategies without separate fitting.

\bibliographystyle{ACM-Reference-Format}
\bibliography{reference}
 
\end{document}